\documentclass[draft, twoside]{article}

\usepackage{amsfonts}
\usepackage{graphicx}
\usepackage{type1cm}

\newcommand{\ket}[1]	{\left| #1 \right\rangle}

\newcommand{\+}         {\dagger}
\newcommand{\eend}      {\hspace{\stretch{1}}\rule{1ex}{1ex}}

\newtheorem{proposition}{Proposition}[section]
\newtheorem{proof}{Proof}[section]
\newtheorem{definition}{Definition}[section]

\newtheorem{algorithm}{Algorithm}[section]
\newtheorem{remark}{Remark}[section]

\begin{document}

\title{Quantum convolutional codes: fundamentals}

\author{H. {Ollivier}\thanks{INRIA, Projet CODES, BP 105, Domaine de
Voluceau, F-78153, Le Chesnay, France}~ and J.-P. Tillich\raisebox{0.5ex}{\scriptsize *}}

\maketitle

\begin{abstract}
  We describe the theory of quantum convolutional error correcting
  codes. These codes are aimed at protecting a flow of quantum
  information over long distance communication. They are largely
  inspired by their classical analogs which are used in similar
  circumstances in classical communication. In this article, we
  provide an efficient polynomial formalism for describing their
  stabilizer group, derive an on-line encoding circuit with linear
  gate complexity and study error propagation together with the
  existence of on-line decoding. Finally, we provide a maximum
  likelihood error estimation algorithm with linear classical
  complexity for any memoryless channel.
\end{abstract}

\section{Introduction} 
Quantum information science has been developed in the past two decades
as a way to process information more efficiently than with classical
means. It lead to great theoretical advances and to impressive
experimental realizations (see~\cite{Pre98a, NC00a} for a review). The
main results motivating the interest for quantum computation concern
integer factorization~\cite{Sho94a} and unsorted database
search~\cite{Gro97a}. Both contribute to the widely accepted idea that
quantum computers are intrinsically more powerful than their classical
analogs, and justify the ever increasing interest for this new model
of computation.

In parallel to these developments, the difficulty of building quantum
information processing devices has been throughly pointed out: the
quantum world is extremely sensitive to interactions with its
surrounding environment~\cite{Zur91a, Zur02a, Zur03a}. This process,
called decoherence, is responsible for the instability of the fragile
quantum superpositions necessary to obtain a speedup over classical
computation~\cite{JL02a}. In absence of any control over the
decoherence process, these quantum devices would be turned into---at
best---classical computers. Fortunately, the discovery of quantum
error correction schemes~\cite{Sho95a}, together with their
fault-tolerant implementation~\cite{Got98a} cleared the future of
quantum computation: quantum codes protect from unwanted evolutions
and noise, whereas their fault-tolerant implementation guarantees
that, below a certain error rate, quantum information processing can
be done without loss of coherence~\cite{DS96a, Zal96a, Got98a, AB99b}.

However, generic encoded quantum computation requires a large overhead
in costly quantum resources: up to now, only a single encoded qubit
has been produced and manipulated successfully on an experimental
quantum processing unit~\cite{KLMN01a}. On the other hand, quantum
communication protocols---e.g.\ quantum key distribution---achieve the
production of large numbers of qubits often represented by some
degrees of freedom of light modes. In most of the protocols, the
manipulation of quantum bits is very limited and errors occur mainly
during the transmission---loss of photons, noise, etc. In the
perspective of quantum communication, we develop a theory of quantum
convolutional error correcting codes. These codes are largely inspired
by their classical analogs~\cite{Lee97a, JZ99a} and share many of
their properties: efficient encoding and decoding circuits and an
efficient maximum likelihood error estimation procedure for any
memoryless channel. As in the classical context, these codes can deal
with infinitely long streams of ``to-be-protected'' information
without introducing unacceptable delays in the transmission.

The article is organized as follows: sec.~\ref{sec:structure}
describes the structure of quantum convolutional codes and introduces
an appropriate formalism; sec.~\ref{sec:encoding} provides an encoding
circuit for this class of codes; sec.~\ref{sec:catastrophicity}
studies error propagation properties, and sec.~\ref{sec:correction}
details the efficient maximum likelihood error estimation
algorithm. Throughout the text, abstract concepts are readily applied
to a previously introduced example of quantum convolutional
code~\cite{OT03a}.

\section{Structure of quantum convolutional codes}\label{sec:structure}

All error protection strategies share many common ingredients. First,
they must define the structure in which quantum information will be
stored and, as second step, explain how information can be manipulated
within this structure. Quantum error correcting codes impose the
information to be stored in a subspace of the total Hilbert space of
the physical qubits. This subspace, $\mathcal C$, is called the code
subspace. $\mathcal C$ is usually further decomposed into---e.g.\
single qubit---subspaces for which elementary operations are then
provided.

However, to arrive at a practical definition of a quantum error
correction scheme, it is usually necessary to further restrict the
possibilities offered by the above general program. One such
restriction leads to stabilizer codes. Those are often compared to
classical linear codes: they are defined by a set of linear
equations---called syndromes---which allow an efficient description of
$\mathcal C$ together with a great flexibility in their design. To
facilitate the introduction of quantum convolutional codes, we will
use abundantly the stabilizer formalism, even though convolutional
codes can be generalized to a wider framework\footnote{In particular,
our main theorem concerning error propagation in
sec.~\ref{sec:catastrophicity} does not rely on the stabilizer
formalism.}.

More precisely, the code subspace $\mathcal C$ of any stabilizer code
is defined as the largest subspace stabilized by an Abelian group $S$
acting on the $N$ physical qubits of the code. In practice, $S$ is a
subgroup of the multiplicative Pauli group $G_N = \mathrm{sp}\,
\{I,X,Y,Z\}^{\otimes N}$, where $I$, $X$, $Y$, $Z$ are the well known
Pauli matrices\footnote{ $ I = \left (\begin{array}{cc} 1 & 0 \\ 0 & 1
\end{array}\right),\; X = \left (\begin{array}{cc} 0 & 1 \\ 1 & 0
\end{array}\right),\; Y = \left (\begin{array}{cc} 0 & -i \\ i & 0
\end{array}\right),\; Z = \left (\begin{array}{cc} 1 & 0 \\ 0 &
-1\end{array}\right).$}. The description of $\mathcal C$ is further
  simplified by the introduction of a set of independent generators
  $\{M_i\}$ of $S$. This leads to the definition of $\mathcal C$ in
  terms of syndromes:
\begin{equation}
\forall\, i, \ \ket \psi = M_i \ket \psi \ \Leftrightarrow \ \ket \psi
\in \mathcal C.
\end{equation}

\subsection{Definition}
The particularity of convolutional codes is to impose a specific form
to the generators of the stabilizer group such that on-line encoding,
decoding and error correction become possible even in the presence of
an infinitely long to-be-protected stream of information. 

However, convolutional codes do not consider groups of qubits
independently of each other: the encoding operation cannot be
decomposed into a tensor product of encoding operations acting on a
small number of qubits. By contrast, an $(N,K)$-block code can protect
such stream only by cutting it into successive $K$-qubit blocks. As a
result, the code subspace defined by these independent applications
can be decomposed as a tensor product of the $N$-qubit subspaces of
each output block. Furthermore, increasing the parameter $K$ is
usually not an option as it requires, in most cases, a quadratic
overhead in the complexity of the encoding circuit~\cite{Got97a} and,
more dramatically, an exponentially growing complexity of the error
estimation algorithm\footnote{This holds for random codes without
particular structure---not belonging to a restricted class--- and with
constant rate as $K$ increases.}.

Quantum convolutional codes are especially designed to offer an
alternative to small block codes in counteracting the effect of
decoherence and noise over long-distance communications while using a
limited overhead of costly quantum resources.

\begin{definition}[$(n,k,m)$-convolutional code]\label{def:definition}
The stabilizer group, $S$, for an $(n,k,m)$-convolutional code is given
by:
\begin{equation}
S = \mathrm{sp}\, \{M_{j,i} = I^{\otimes j\times n} \otimes M_{0,i}, \
1\leq i \leq n-k, \ 0 \leq j\}, \label{eq:definition}
\end{equation}
where $M_{0,i} \in G_{n+m}$. Above $M_{j,i}$'s are required to be
independent and to commute with each other.
\end{definition}

\begin{remark}\rm
As expected, the length of the code (i.e. the number $N$ of physical
qubits of the code) as well as the number of logical qubits are left
unspecified. In fact, the maximum value of the integer $j$ controls
this length implicitly. However, and contrarily to block codes, this
maximum value does not need to be known in advance for encoding and
decoding qubits. Instead, it will be fixed a posteriori when the
transmission ends. This specific issue will be addressed in
sec.~\ref{sec:encoding}. Hence, in most situations the length of the
code can simply be set to infinity. The only associated restriction is
to consider operators whose support\footnote{In this article the
definition of support of an element $A$ of the Pauli group is---rather
unconventionally---the smallest block of consecutive qubits on which
$A$ acts non-trivially.\label{foot:support}} has size of order 1. This
also explains why in Eq.~(\ref{eq:definition}) the $M_{j,i}$'s seem to
have different length: in the rest of the article we simply assume
that the operators are ``padded'' by identities on the right-most
physical qubits to adjust them to the appropriate length.\eend
\end{remark}

With this remark in mind, the structure of the stabilizer group
generators can be summarized easily with the help of a semi-infinite
matrix $M$:
\begin{equation}
M = \left(
\begin{picture}(170,65)

\put(0,15){\framebox(70,47){$\begin{array}{c} M_{0,1} \\ \vdots \\ M_{0,n-k}\end{array}$}}
\put(50,-35){\framebox(70,47){$\begin{array}{c} M_{1,1} \\ \vdots \\ \quad M_{1,n-k}\end{array}$}}

\put(130,-55){\makebox(0,0)[lb]{$\ddots$}}

\multiput(70,-35)(0,4){12}{\line(0,1){2}}

\put(45,-20){\vector(0,1){30}}
\put(45,00){\vector(0,-1){30}}
\put(43,-26){\rotatebox{90.0}{\makebox(0,0)[lb]{$n-k$}}}

\put(55,-40){\vector(1,0){15}}
\put(65,-40){\vector(-1,0){15}}
\put(57,-48){\makebox(0,0)[lb]{$m$}}

\put(75,-40){\vector(1,0){45}}
\put(115,-40){\vector(-1,0){45}}
\put(93,-48){\makebox(0,0)[lb]{$n$}}

\end{picture}
\right) \label{eq:general_structure}
\end{equation}
Each line of the matrix represents one of the $M_{j,i}$ and each
column a different qubit. A given entry in $M$ is thus the Pauli
matrix for the corresponding qubit and generator.  The rectangles
represent graphically which qubits are potentially affected by the
action of the generators. The form of Eq.~(\ref{eq:general_structure})
visually emphasizes the structure of convolutional codes: 
\begin{itemize}
\item $M$ has a block-band structure;
\item the overlap of $m$ qubits between two neighboring sets of
generators forces to consider the code subspace as a whole. 
\end{itemize}
By contrast, for a block code used repeatedly to protect an infinitely
long stream of qubits, the above parameter $m$ would be equal to 0.

\begin{remark}\rm
In addition to the above generators, and in order to properly account
for the finiteness of real-life communications, a few other generators
will be added to the matrix $M$. This will however not interfere with
the rest of this section.\eend
\end{remark}

\subsection{Polynomial representation}
Although, it is in principle possible to carry out a complete the
analysis of the code with the matrix $M$ only, we will introduce a
polynomial formalism which greatly simplifies this task. Such
formalism is the exact translation of the polynomial formalism for
classical convolutional codes. Its advantage is to capture in a
convenient and efficient way the fact that the generators in $M$ are
$n$-qubit shifted versions of the $M_{0,i}$'s.

More precisely, for a $(n,k,m)$-convolutional code, we define the
delay operator $D$ acting on any element $A$ of the Pauli group of the
physical qubits with bounded
support\raisebox{0.85ex}{\scriptsize\ref{foot:support}} by:
\begin{equation}
D[A] = I^{\otimes n} \otimes A,
\end{equation}
with the same ``padding rule'' as before. Naturally, one can consider
powers of $D$ as repeated applications of the delay operator. For
instance, the generators of the code can now be written as:
\begin{equation}
M_{j,i} = D^j[M_{0,i}], \ 0 \leq j, \ 1 \leq i \leq n-k.
\end{equation}
Therefore, and to further continue with simplifications, it is
obviously not necessary to keep more than the first $n-k$ lines of the
matrix $M$ defined in Eq.~(\ref{eq:general_structure}). All the
omitted ones can be easily recovered by applying $D$ the appropriate
number of time.

In addition to applying a single $D^j$ to an element of the Pauli
group, it is, under certain conditions, possible to consider more
complex operations---for instance, these will be necessary for
deriving the encoding circuit. Namely, consider $A$, an element of the
Pauli group with bounded support, such that $A$ and $D^j[A]$ commute
for any value of $j$. Then, the full polynomial ring $GF_2[D]$ can act
on $A$. For $P(D) = \sum_j \alpha_j D^j$, the action of $P(D)$ on $A$
is naturally defined as:
\begin{equation}
P(D)[A] = \prod_j \alpha_j D^j[A].
\end{equation}
Above, the commutation relation is crucial: the sum operation in
$GF_2[D]$ is commutative and must therefore be translated into another
commutative operation---here the product---on the multiplicative group
spanned by $\{D^j[A]\}_j$. 

Finally, we will sometimes use a short hand in our notation and,
instead of restricting ourselves only to polynomials in $D$, consider
formal Laurent series acting on $A$. In such case, we do not really
need to define the action of negative powers of $D$, but we impose
that, at the end of the calculation---possibly concerning several
operators---, all the negative powers of $D$ are removed by
globally\footnote{This means on all the operators involved in the
calculation.} applying the smallest possible positive power of
$D$. For instance, if we end with
\begin{equation}
L(D)[A] = \left(\sum_{j=-p}^q \alpha_j D^j \right) [A], 
\end{equation} 
it will be turned into  
\begin{eqnarray}
P(D)[A] & = & \left(D^p \sum_{j=-p}^q \alpha_j D^j \right) [A]
\nonumber \\ 
& = & \left(\sum_{j=-p}^q \alpha_j D^{j+p} \right) [A].
\end{eqnarray}

In practice, the representation of the code generators as a matrix $M$
with entries $I$, $X$, $Y$, $Z$ is often replaced by the one
of~\cite{CRSS97a}. In this representation, the first $n-k$ generators
of an $(n,k,m)$-convolutional code would be written as a pair of
$(n+m)\times(n-k)$ binary matrices arranged side by side\footnote{This
representation as a pair of binary vectors or matrices is not
restricted to elements of the stabilizer group, and can indeed be used
for any element of $G_{n+m}$.}. Each line corresponds to a generator
and each column to a qubit. A 1 for the left matrix indicates the
presence of an $X$ or $Y$ and, similarly, a 1 for the right matrix
indicates the presence of a $Y$ or $Z$. Within this framework, it is
easy to realize that the polynomial formalism can be fruitfully
extended to lead an even more compact notation for the generators of
the stabilizer group.

First, recall that the addition of two pairs of binary vectors simply
results in the multiplication of the corresponding generators provided
that these commute. For instance, suppose $A$ and $B$ are two elements
of $G_n$, and $(A_X|A_Z)$, $(B_X|B_Z)$ their respective
representations as pair of binary vectors. In such case, the operator
$A\otimes B$ is represented by $(A_X:B_X| A_Z:B_Z)$ where ``:''
indicates the concatenation of the vectors. With the polynomial
formalism, we also have $A\otimes B = A\times D[B]$, which leads
to\footnote{Here again we apply the implicit ``padding rule'' to
adjust the length of the vectors.} $(A_X:B_X| A_Z:B_Z) = (A_X|A_Z) +
D[(B_X|B_Z)]$. Here, the commutation of $A$ and $D[B]$ is trivially
verified since their supports do not intersect. This last equality
suggests the following modification of the representation. A generic
element $P$ of the Pauli group of the physical qubits with bounded
support is represented by a pair of length $n$ vectors with
coefficients in $GF_2[D]$ such that, by definition,
\begin{eqnarray}
(P_X|P_Z) & = & (P_X^{(0)}:P_X^{(1)}:P_X^{(2)}:\ldots |
P_Z^{(0)}:P_Z^{(1)}:\ldots)\nonumber 
\\ & = & (P_X^{(0)} + D\times P_X^{(1)} + D^2\times P_X^{(2)} + \ldots |
P_Z^{(0)} + D\times P_Z^{(1)} + \ldots),
\end{eqnarray} 
where the $P_X^{(j)}$'s and $P_Z^{(j)}$'s are length $n$ binary
vectors.

All these new concepts are best illustrated by applying them to a
particular convolutional code. The simplest one with non-trivial
behavior is the $(5,1,2)$-convolutional code given in~\cite{OT03a}:
\begin{equation}
\begin{array}{lcp{1em}@{}p{1em}@{}p{1em}@{}p{1em}@{}p{1em}@{}p{1em}@{}p{2em}} 
M_{0,1} & = & $Z$&$X$&$X$&$Z$&$I$&$I$&$I$, \\ 
M_{0,2} & = & $I$&$Z$&$X$&$X$&$Z$&$I$&$I$, \\ 
M_{0,3} & = & $I$&$I$&$Z$&$X$&$X$&$Z$&$I$, \\
M_{0,4} & = & $I$&$I$&$I$&$Z$&$X$&$X$&$Z$, \\
M_{j,i} & = & \multicolumn{7}{l}{D^{j}[M_{0,i}], \ 0 \leq j.}
\end{array}
\end{equation}
Thus, using the pair of polynomial matrices representation, the
generators of the stabilizer group can be written:
\begin{equation}
M = \left(
\begin{array}{p{1em}p{1em}p{1em}p{1em}p{1em}|p{1em}p{1em}p{1em}p{1em}p{1em}}
0&1&1&0&0 & 1&0&0&1&0 \\
0&0&1&1&0 & 0&1&0&0&1 \\
0&0&0&1&1 & $D$&0&1&0&0 \\
$D$&0&0&0&1 & 0&$D$&0&1&0
\end{array}
\right).\label{eq:polynomial_matrix}
\end{equation}

\subsection{Generalized commutator}
In the context of block codes, the main reason justifying the
introduction of the representation of the elements of the Pauli group
as pairs of binary vectors~\cite{CRSS97a, Got97a} is the existence of
an easy way to compute the group commutator. We shall see below that
the same kind of advantage holds for the representation as pair of
polynomial vectors.

First, consider two elements $A = (A_X | A_Z)$ and $B = (B_X|B_Z)$ of
$G_N$. It is then easy to check on their representation as pair of
binary vectors that,
\begin{equation}
AB = BA \ \Leftrightarrow \ A_X B_Z + A_Z B_X = 0,
\end{equation}
where we use the standard inner product of two vectors of length $N$
and addition modulo 2.

Now suppose that $P$ and $Q$ are two elements of the Pauli group of
the physical qubits of an $(n,k,m)$-convolutional code, and $(P_X(D)|
P_Z(D))$, $(Q_X(D)|Q_Z(D))$ their representation as pair of polynomial
vectors. Using the above method, one can conclude that the commutation
of $P$ and $Q$ is simply expressed by:
\begin{equation}
PQ = QP \ \Leftrightarrow \ \sum_l P_X^{(l)} Q_Z^{(l)} + P_Z^{(l)}
Q_X^{(l)} = 0,
\end{equation}
where $P_X(D) = \sum_j P_X^{(j)} D^j$ with $P_X^{(j)}$ a binary vector
of length $n$ and similarly for $P_Z(D)$, $Q_X(D)$, $Q_Z(D)$. This
also leads to,
\begin{equation}
D^r[P]D^s[Q] = D^s[Q]D^r[P] \ \Leftrightarrow \ \sum_l P_X^{(l+s)}
Q_Z^{(l+r)} + P_Z^{(l+s)} Q_X^{(l+r)} = 0.
\end{equation}
The last equation is particularly interesting since its right hand
side is the coefficient of $D^{s-r}$ in $P_X(D)Q_Z(1/D) +
P_Z(D)Q(1/D)$. Therefore, one can readily conclude that the
representation as pair of polynomial vectors allows an easy
computation of the ``generalized commutation relation''---i.e.\ the
commutation of any $n$-qubit shifted version of $P$ with any $n$-qubit
shifted version of $Q$---:
\begin{eqnarray}
& \forall\, r,s, \ D^r[P] D^s[Q] = D^s[Q] D^r[Q] & \nonumber \\ 
& \Leftrightarrow & \label{eq:general_commutation}\\
&  P_X(D)Q_Z(1/D) + P_Z(D)Q_X(1/D) = 0.\nonumber&
\end{eqnarray}

We will see below that this property of the polynomial representation
is crucial as it allows the derivation of almost all the encoded Pauli
operators by considering only the first $n-k$ generators $M_{0,i}$'s
of the stabilizer group.

\subsection{Encoded Pauli operators}
The encoded Pauli operators for a quantum error correcting code are
some operators of the Pauli group of the physical qubits which allow
the manipulation of the information without requiring any
decoding. More precisely, these are operators that leave the code
subspace $\mathcal C$ globally invariant, but which have a non-trivial
action on it. Indeed, it is possible to require such operators to
reproduce exactly the commutation relations of the Pauli group for the
encoded qubits. This is mathematically expressed
by~\cite{Got97a}\footnote{In all this article, and following the
notation of~\cite{Got97a}, the encoded Pauli operators are denoted by,
e.g.\ $\overline X$ and $\overline Z$.}:
\begin{eqnarray}
\overline X_i, \ \overline Z_i & \in & N(S)/S, \label{eq:pauli_constraint1}\\
{[}\overline X_i, \overline X_j{]} & = & 0, \label{eq:pauli_constraint2}\\
{[}\overline Z_i, \overline Z_j{]} & = & 0, \label{eq:pauli_constraint3}\\
{[}\overline X_i, \overline Z_j{]} & = & 0, \ i \neq j, \label{eq:pauli_constraint4} \\
\{ \overline X_i, \overline Z_i \} & = & 0, \label{eq:pauli_constraint5}
\end{eqnarray}
where the index $i$ in $\overline X_i$ and $\overline Z_i$ denotes the
$i$-th logical qubit.

In the rest of this paragraph we exploit
Eq.~(\ref{eq:general_commutation}) to find an algorithmic procedure
for deriving the $\overline X_i$'s and $\overline Z_i$'s. First we
define the standard polynomial form of $M$ and, as a second step, we
translate
Eqs.~(\ref{eq:pauli_constraint1}--\ref{eq:pauli_constraint5}) into a
set of equations for polynomial vectors which can be solved easily.

To obtain the standard polynomial form for the generators of the
stabilizer group one can perform two Gaussian
eliminations\footnote{See also~\cite{Got97a} for a similar procedure
for block codes} on $M$ written in its representation as pair of
polynomial matrices over $GF_2[D]$. This can be done by using line
additions, column swaps and multiplication of a line by a power of
$D$:
\vspace{2.5ex}
\begin{equation}
M_{\mathrm{std}} =
\left( \begin{array}{ccc|ccc}
\raisebox{0ex}[1.5ex]{$\overbrace{A(D)}^r$} & 
\raisebox{0ex}[1.5ex]{$\overbrace{B(D)}^{n-k-r}$} & 
\raisebox{0ex}[1.5ex]{$\overbrace{C(D)}^k$} & 
\raisebox{0ex}[1.5ex]{$\overbrace{E(D)}^r$} & 
\raisebox{0ex}[1.5ex]{$\overbrace{F(D)}^{n-k-r}$} & 
\raisebox{0ex}[1.5ex]{$\overbrace{G(D)}^k$} \\
0 & 0 & 0 & J(D) & K(D) & L(D)
\end{array} \right) 
\!\!\!\!
\begin{array}{l} \}r \\ \}n-k-r \end{array} \label{eq:standardD}
\end{equation}
where $A(D)$ and $K(D)$ are diagonal matrices with polynomial
coefficients, and where $r$ is the rank of the $X$-part of $M$. 

By definition, $A(D)$ has full rank. In fact, this holds for $K(D)$ as
well: if it was not the case, then there would exist a line with
zeroes everywhere except for at least one position in the first $r$
columns of the $Z$-part. Then, the operator corresponding to this line
cannot commute in the generalized sense with all the other generators,
which would contradict the assumption that the stabilizer group $S$
generated by $M_{\mathrm{std}}$ is Abelian.

We now turn to the determination of the encoded Pauli operators. Here,
we restrict our search to operators that preserve the convolutional
nature of the code: we want to find a finite set of independent
operators with bounded support which generate through $n$-qubit
shifts---almost all---the encoded Pauli operators\footnote{For the
purpose of introducing the theory of quantum convolutional codes, it
is not necessary to consider encoded Pauli operators that do not
respect the convolutional structure of the code. However, in more
elaborated error correction scheme, this might prove to be
useful.}. This can be accomplished by considering a $k$-line matrix,
\begin{equation}
\overline X = (U_1(D),U_2(D),U_3(D)|V_1(D),V_2(D),V_3(D)),
\end{equation}
representing the encoded $\overline X$ operators---the rest of the
discussion shows that such encoded Pauli operators exist. Since these
operators can be multiplied by any element of the stabilizer group,
$U_1(D)$ and $V_2(D)$ can be set to 0. The generalized commutation
with the lines of $M$ imposed by Eq.~(\ref{eq:pauli_constraint1}) can
be simply written:
\begin{eqnarray}
&&\left( \begin{array}{ccc|ccc} A(D) & B(D) & C(D) & E(D) & F(D) &
G(D) \\ 0 & 0 & 0 & J(D) & K(D) & L(D)
\end{array} \right)
\left( \begin{array}{c}
V_1^T(1/D)\\ 0 \\ V_3^T(1/D) \\ \hline 0 \\ U_2^T(1/D) \\ U_3^T(1/D)
\end{array}\right) = \nonumber \\
&& \qquad = \left( \begin{array}{c} 
0 \\ 0 
\end{array}\right).\label{eq:Xbar_commute}
\end{eqnarray}
On the other hand, Eq.~(\ref{eq:pauli_constraint2}) is expressed by
\begin{equation}
U_3(D)V_3^T(1/D) + V_3(D)U_3^T(1/D) = 0,
\end{equation}
which can be trivially satisfied with $V_3(D) = 0$ and $U_3(D) =
\Lambda(D)\times I$, where $\Lambda(D)$ is a non-zero polynomial of
$GF_2[D]$. This choice guarantees that the operators in $\overline X$
together with their $n$-qubit shifted versions are independent of each
other and from the generators of $S$. In this case,
Eq.~(\ref{eq:Xbar_commute}) becomes,
\begin{equation}
\left( \begin{array}{c}
A(D)V_1(1/D)^T + F(D)U_2(1/D)^T + G(D)U_3(1/D)^T \\
K(D)U_2(1/D)^T + L(D)U_3(1/D)^T
\end{array}\right) = 
\left( \begin{array}{c} 
0 \\ 0 
\end{array}\right).
\end{equation}

Then we can write the encoded $\overline X$ operators:
\begin{eqnarray}
U_1(D) &=& 0 \label{eq:U1}\\
U_2(D) &=& L^T(1/D) K^{-1}(1/D) \Lambda(D) \\ 
U_3(D) &=& \Lambda(D)\times I \\
V_1(D) &=& \left(U_2(D) F(1/D)^T + \Lambda(D) G(1/D)^T\right) A^{-1}(1/D) \\
V_2(D) &=& 0 \\
V_3(D) &=& 0.\label{eq:V3}
\end{eqnarray}
One must realize that the encoded Pauli operators $\overline X$ are
not yet properly defined as the division by polynomials is in general
problematic. The reason is that generic polynomial fractions cannot be
written as finite formal Laurent series. Thus, the operators that they
describe have an unbounded support. In such case, and without further
modifications, the formalism introduced earlier imposes transmissions
of infinite length. However, when the result of the division
can be written with a finite Laurent series, such operation is
permitted.

\begin{definition}[Conditioning polynomial]
The conditioning polynomial $\Lambda(D)$ of a convolutional
code is the non-zero polynomial with minimum degree such that the
equations (\ref{eq:U1}--\ref{eq:V3}) only involve finite Laurent
series.
\end{definition}
As it can be seen easily, the conditional polynomial always exists,
and the $\overline X$'s operators are well defined. They correspond to
operators with a finite support, respecting the convolutional
structure of the code.

We now turn to the derivation of some $\overline Z$'s by applying the
same tools. First note that once the $\overline X$'s are fixed, there
is a unique set of valid $\overline Z$'s. Quite surprisingly, we will
also see here that it is not always possible to impose to the
$\overline Z$'s the convolutional structure---the invariance by
$n$-qubit shifts. 

For instance, first define the $k$-line matrix
\begin{equation}
\overline Z = (0,U_2'(D),U_3'(D)|V_1'(D),0 ,V_3'(D)).
\end{equation}
Above, the zeroes have been set for the same reason as in the
derivation of the $\overline X$'s. In addition to satisfying an
equation similar to Eq.~(\ref{eq:Xbar_commute}), the matrix $\overline
Z$ must anti-commute in the generalized sense with $\overline X$,
Eq.~(\ref{eq:pauli_constraint5}). Equivalently, this can be expressed
as $V_3'(D) U_3(1/D)^T = I$, which can be fulfilled if and only if
$V_3' = I / \Lambda(D)$. As discussed above, only when $\Lambda(D)$ is
a monomial in $D$ does $V_3'(D)$ correspond to a valid polynomial
vector (i.e.\ $1/\Lambda(D)$ is a bounded Laurent series). In
this latter case, we obtain $\overline Z$:
\begin{eqnarray}
U_1'(D) &=& 0 \\ 
U_2'(D) &=& 0 \\ 
U_3'(D) &=& 0 \\ 
V_1'(D) &=& C^T(1/D) A(1/D)^{-1} / \Lambda(1/D) \label{eq:what_name}\\
V_2'(D) &=& 0 \\ 
V_3'(D) &=& I / \Lambda(1/D).
\end{eqnarray}
Note that for $\Lambda(D)$ to be a monomial, all the $A_{i,i}(D)$'s
must be monomials as well, so that Eq.~(\ref{eq:what_name}) is
automatically well defined.

\begin{remark}\rm
The obvious question raised by this derivation concerns the case where
$\Lambda(D)$ is not a monomial. The rigorous answer will be given in
sec.~\ref{sec:catastrophicity} where it will be shown that if such
code were to be used, it would have bad error propagation
properties.\footnote{Only the $\overline X$ operators are used to
derive the encoding circuit. Then, if one renounces to manipulate
information in its encoded form, the code can be, in principle,
successfully used to protect quantum information.} One can also
consider the following hand-waving argument: when $\Lambda(D)$ is not
a monomial, and for a finite length communication, the $\overline Z$'s
have a support with a size of the order of the length of the
code. Thus, if one implements an encoded phase flip by applying
individual $Z$'s on the physical qubits with finite precision, then
for long streams of to-be-protected information this will result in an
error with probability close to 1.\eend
\end{remark}

Finally, and to conclude this section on the structure of
convolutional codes, we should count how many logical qubits are
described by our construction in the case of a finite transmission. To
simplify this discussion, we define the integer $\lambda$ as the
highest degree in the polynomial matrices $\overline X$ and $\overline
Z$.  This sets an upper bound on the size of the support of any of the
$\overline X$'s and $\overline Z$'s: they extend on at most
$\lambda+1$ consecutive $n$-qubit blocks. Further consider the
stabilizer group $S$ generated by the $\{M_{j,i}\}$ for $0 < i \leq
n-k$ and $0 \leq j < p$ with $p > \lambda$. In this case, the above
derivation leads to at least\footnote{Here, we consider an integer
number of physical $n$-qubit blocks. I wrote ``at least'' because it
is possible that the support of some of the $\overline X$ and
$\overline Z$ is smaller than $n\times (\lambda+1)$.}  $k\times (p +
\lceil m/n \rceil - \lambda)$ logical qubits while we used
$(n-k)\times p$ generators and $(p + \lceil m/n \rceil) \times n$
physical qubits. Therefore, only $\lceil m/n \rceil \times (n-k) +
\lambda k $ logical qubits---a number independent of $p$---do not
follow the convolutional structure of the code. These will simply be
discarded in the encoding process as this does not change the
asymptotic rate of the code. This can be done consistently with the
stabilizer formalism by adding their encoded $\overline Z$ operators
to the generators of $S$.

By working out the example of Eq.~(\ref{eq:polynomial_matrix}), one
easily finds the standard form of $M$,
\begin{equation}
M_{\mathrm{std}} = \left(
\begin{array}{p{1em}p{1em}p{1em}p{1em}p{1em}|p{3em}p{1em}p{1em}p{1em}p{1em}}
$D$& 0 & 0 & 0 & 1 &   0  &$D$& 0 & 1 & 0 \\
0  & 1 & 0 & 0 & 1 & $1+D$& 1 & 1 & 1 & 1 \\
0  & 0 & 1 & 0 & 1 & $D$  & 1 & 1 & 0 & 1 \\
0  & 0 & 0 & 1 & 1 & $D$  & 0 & 1 & 0 & 0 \\
\end{array}
\right).
\end{equation}
The $\overline X$ operators are obtained from a single 5-dimensional
vector, with the polynomial $\Lambda(D)$ equal to $1$:
\begin{equation}
\begin{array}{lcl}
\overline X & = & (0, 0, 0, 0, 1 | 0, 1, 1, 0, 0), \\
\overline Z & = & (0, 0, 0, 0, 0 | D, 1, 1, 1, 1).
\end{array}
\end{equation}

\section{Encoding}\label{sec:encoding}
This section provides an operational method to arrive at an encoding
circuit which respects the convolutional structure of the code: a
simple unitary operation---inde\-pen\-dent of the length of the
to-be-protected stream---and its $n$-qubit shifted versions will be
applied successively to arrive at the protected state. Therefore, the
complexity of this scheme in terms of number of gates in the encoding
circuit only grows linearly with the number of encoded qubits. This is
of particular relevance since dealing with convolutional codes as if
they were generic block codes would lead to an encoding circuit with
quadratic gate complexity. It would also require increasing precision
in the applications of the encoding gates and would cause severe
delays in the transmission of the information.

The derivation of the encoding circuit will nonetheless be very
similar to the one for block codes~\cite{Got97a}. Here, instead of the
usual standard form for the generators, we use the standard polynomial
form. The circuit that will be obtained is relative to the encoding of
$q\times k$ logical qubits. The encoded Pauli operators corresponding
to these qubits will be denoted $\overline X_{j,i}$ and $\overline
Z_{j,i}$. For instance, $\overline X_{0,i}$ is defined as the $i$-th
line of the $\overline X$ matrix derived in the previous section,
$\overline X_{j,i} = D^j[\overline X_{0,j}]$, and similarly for the
$\overline Z$'s.

The encoding circuit maps the to-be-protected qubits $c_{j,i}$ onto
the code subspace. Its action on the computational basis can be
written as:
\begin{equation}
\ket {c_{0,1},\ldots, c_{q-1, k}} \rightarrow
\left(\prod_{i,j}\frac{1+M_{j,i}}{\sqrt 2} \right) \prod_{r,s}
\overline X_{s,r}^{c_{s,r}} \ket {0,\ldots,
0},\label{eq:encoding_map}
\end{equation}
for $c_{s,r} \in \{0,1\}$, $0<i\leq n-k$, $0\leq j <q+\lambda$, $0\leq
s<q$ and $1\leq r \leq k$.\footnote{Here $\lambda$ is defined as the
in the previous section. With this definition, the operators
$\overline X$ have support on at most $\lambda + 1$ consecutive
$n$-qubit blocks. The choice $j < q+\lambda$ then ensures that the
support of each logical qubit is covered by the same number of
generators of the stabilizer group.} This operation can be decomposed
in two steps. The first one, $\prod_{r,s} \overline
X_{s,r}^{c_{s,r}}$, applies the different flip operators depending on
the value of the to-be-protected qubits in the computational
basis. The second, $\prod_{i,j}({1+M_{j,i}})/{\sqrt 2}$, projects this
state onto the code subspace.\footnote{The way of writing this
projection follows from the realization that any element of the
stabilizer group is a product where each generator appears at most
once---any element of the Pauli group is its own inverse.}

We first focus on the conditional application of the $\overline X$'s:
\begin{equation}
\ket {c_{0,1},\ldots, c_{q-1, k}} \rightarrow \prod_{r,s} \overline
X_{s,r}^{c_{s,r}} \ket {0,\ldots, 0}.\label{eq:pauli_X}
\end{equation}
The number of $n$-qubit blocks involved in the right hand side of
Eq.~(\ref{eq:encoding_map}) is equal to $q+\lambda + \lceil m/n
\rceil$. Hence, the first requirement is to supplement the
to-be-protected stream of information with some ancillary qubits
prepared in the $\ket 0$ state. Both are arranged in the following
way:
\begin{equation}
\begin{array}{l}
\ket{c_{0,1},\ldots, c_{q-1, k}} \rightarrow \\
\quad |
\raisebox{0ex}[5.5ex]{$\overbrace{0\ldots 0}^{n\times\lambda-k}$},
c_{0,1}\ldots c_{0, k}, 
\raisebox{0ex}[1.5ex]{$\overbrace{0\ldots 0}^{n-k}$}, 
c_{1, 0} \ldots c_{1, k}, 
\raisebox{0ex}[1.5ex]{$\overbrace{0\ldots 0}^{n-k}$}, 
c_{q-1,k} \ldots c_{q-1,k},
\raisebox{0ex}[1.5ex]{$\overbrace{0\ldots 0}^{\lceil m/n \rceil \times n}$}
\rangle.
\end{array}\label{eq:stream_ancillas}
\end{equation}
The notation $\overline X_{s,r}^{c_{s,r}}$ means that $\overline
X_{s,r}$ needs to be applied on the all-zeroes state if and only if
$c_{s,r} = 1$. Now, in the standard polynomial form, $\overline
X_{s,r}$ has a factor $X$ exactly at the position of $c_{s,r}$ in the
state of the right hand side of
Eq.~(\ref{eq:stream_ancillas}). Therefore, if all the other logical
qubits are set to zero, the output state of Eq.~(\ref{eq:pauli_X}) can
be obtained from the right hand side of Eq.~(\ref{eq:stream_ancillas})
by applying $\overline X_{s,r}$---without the above mentioned
$X$---conditioned on qubit $c_{s,r}$. Unlike for quantum bock codes,
these conditional operations can confuse each other when the
conditioning polynomial $\Lambda(D)$ (see sec.~\ref{sec:structure}) is
not a monomial\footnote{Here, for sake of generality, we describe the
encoding circuit without imposing $\Lambda(D)$ to be a monomial even
though in this case, the encoding shows bad error propagation
properties.}. In this situation, applying $\overline X_{s,r}$ might
flip some control qubits $c_{s',r}$ for $s' < s$. Therefore, these
modified qubits $c_{s',r}$ cannot be used anymore to condition the
application of $\overline X_{s',r}$. This also indicates the way-out
of this problem: when the $\overline X$'s are applied by increasing
successively the index $s$ by one, there is no risk that one
application flips a qubit later used to condition another $\overline
X$.

\begin{figure}[tp]
\begin{center}
\begin{picture}(70,150)

\multiput(-20,150)(0,-10){4}{\makebox{$\ket 0$}}
\multiput(-20,100)(0,-10){4}{\makebox{$\ket 0$}}
\multiput(-20, 50)(0,-10){4}{\makebox{$\ket 0$}}

\put(-20,110){\makebox{$c_{0,1}$}}
\put(-20, 60){\makebox{$c_{1,1}$}}
\put(-20, 10){\makebox{$c_{2,1}$}}

\multiput(0,150)(000,-10){15}{\line(1,0){10}} 

\put(14,110){\circle*{3}}
\put(10,110){\line(1,0){8}}
\put(10,126){\framebox(8,8){Z}}
\put(10,136){\framebox(8,8){Z}}
\put(14,110){\line(0,1){16}}
\put(14,134){\line(0,1){2}}
\put(10,150){\line(1,0){8}}
\put(10,120){\line(1,0){8}}

\put(14,60){\circle*{3}}
\put(10,60){\line(1,0){8}}
\put(10,76){\framebox(8,8){Z}}
\put(10,86){\framebox(8,8){Z}}
\put(14,60){\line(0,1){16}}
\put(14,84){\line(0,1){2}}
\put(10,100){\line(1,0){8}}
\put(10,70){\line(1,0){8}}

\put(14,10){\circle*{3}}
\put(10,10){\line(1,0){8}}
\put(10,26){\framebox(8,8){Z}}
\put(10,36){\framebox(8,8){Z}}
\put(14,10){\line(0,1){16}}
\put(14,24){\line(0,1){2}}
\put(10,50){\line(1,0){8}}
\put(10,20){\line(1,0){8}}

\multiput(18,150)(000,-10){15}{\line(1,0){50}} 

\end{picture}
\caption[Partial encoding circuit for the 5-qubit convolutional code
in the standard polynomial form.]{Circuit for generating the state
$\prod_{r,s} \left(\overline {X_{s,r}}\right)^{c_{s,r}} \ket
{0,\ldots, 0}$ for the 5-qubit convolutional code. For obvious
reasons, the control-$Z$ operations have been kept even though they
act on $\ket 0$ and should be simplified: this part of the encoding
circuit would be reduced to no circuit at all!}\label{fig:x_encoding}
\end{center}
\end{figure}
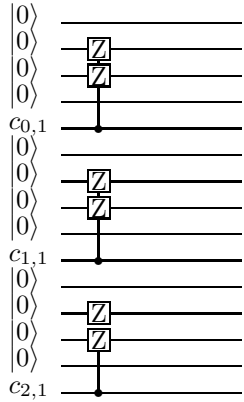

For the example given in Eq.~(\ref{eq:polynomial_matrix}), this part
of the encoding circuit is illustrated in
Fig.~\ref{fig:x_encoding}. 

The rest of the encoding circuit must implement the effect of the
projection onto the code subspace for this partially encoded
state:\footnote{The method described here details how to obtain the
encoding circuit when the generators $M_{j,i}$ have a positive
sign. When this is not the case, the procedure described here must be
modified so that a $Z$ gate is applied to the qubit conditioning the
application of those particular $M_{j,i}$ with a negative sign.}
\begin{equation}
\prod_{r,s} \overline X_{s,r}^{c_{s,r}} \ket {0,\ldots, 0} \rightarrow
\left(\prod_{i,j}\frac{1+M_{j,i}}{\sqrt 2} \right) \prod_{r,s}
\overline X_{s,r}^{c_{s,r}} \ket {0,\ldots, 0}.\label{eq:projection}
\end{equation}
There are two classes of $M_{j,i}$'s. Either $M_{j,i}$ is a tensor
product of $I$'s and $Z$'s only, or there is a polynomial $A_{i,i}$ on
the $i$-th column of the $X$ part when it is expressed in the standard
polynomial form (see sec.~\ref{sec:structure}). In the first case,
nothing needs to be done. In the latter, consider the $i$-th qubit of
the $(j+\mathrm{deg}\, A_{i,i}(D))$-th n-qubit block in
Eq.~(\ref{eq:projection}). The resulting state is an equal weight
superposition of a state with a $\ket 0$ and a state with a $\ket 1$
on the previously mentioned qubit. This can be created by first
applying a Hadamard gate for this qubit, which later controls the
application of $M_{j,i}$---ignoring the $X$ factor for the control. If
there is a $Z$ factor for the control qubit, it does not need to be
conditioned on anything and can be applied right after the Hadamard
gate. Once again, since $A_{i,i}$'s are not required to be monomials,
the above operations might confuse each other when a control qubit,
supposedly still in its initial $\ket 0$ state, has indeed already
been modified. As before, this can be overcome by applying the
conditional gates and increasing the index $j$ one by one successively.

\begin{remark}\rm
For sake of simplicity in the presentation of the whole encoding
circuit, the usual simplifications corresponding to the removal of
control-$Z$ gates acting on a target in state $\ket 0$ have not been
described. Of course, these should be performed to obtain a simpler
circuit.\eend
\end{remark}

\begin{remark}\rm
Note also that the circuit described in this section encodes the
qubits on-line: 
\begin{itemize}
\item the second step rotating the partially encoded state
into the code subspace can start before all the $\overline X$'s are
applied;
\item sending the qubits can be done before all the stream has been encoded.
\end{itemize}
This is a simple consequence of the fact that each conditional gate in
the circuit acts only on the last $\lambda + 1$ $n$-qubit blocks.\eend
\end{remark}

\begin{figure}[tp]
\begin{center}
\begin{picture}(220,170)

\multiput(-20,150)(0,-10){4}{\makebox{$\ket 0$}}
\multiput(-20,100)(0,-10){4}{\makebox{$\ket 0$}}
\multiput(-20, 50)(0,-10){4}{\makebox{$\ket 0$}}
\put(-20,0){\makebox{$\ket 0$}}

\put(-20,110){\makebox{$c_{0,1}$}}
\put(-20, 60){\makebox{$c_{1,1}$}}
\put(-20, 10){\makebox{$c_{2,1}$}}

\multiput(0,150)(000,-10){16}{\line(1,0){10}} 

\multiput(10,136)(000,-10){3}{\framebox(8,8){H}}
\multiput(10, 96)(000,-10){4}{\framebox(8,8){H}}
\multiput(10, 46)(000,-10){4}{\framebox(8,8){H}}
\multiput(10, 110)(000,-50){3}{\line(1,0){8}}
\put(10,150){\line(1,0){10}}
\put(10,-4){\framebox(8,8){H}}

\multiput(18,150)(000,-10){16}{\line(1,0){7}} 

\multiput(25, 150)(000,-50){4}{\line(1,0){8}}
\multiput(25, 136)(000,-50){3}{\framebox(8,8){Z}}
\multiput(25, 126)(000,-50){3}{\framebox(8,8){Z}}
\multiput(25, 120)(000,-50){3}{\line(1,0){8}}
\multiput(25, 110)(000,-50){3}{\line(1,0){8}}

\multiput(33,150)(000,-10){16}{\line(1,0){7}} 

\put(44,100){\circle*{3}}
\put(40,100){\line(1,0){8}}
\put(40,106){\framebox(8,8){X}}
\put(44,100){\line(0,1){6}}
\put(40,120){\line(1,0){8}}
\put(40,130){\line(1,0){8}}
\put(40,140){\line(1,0){8}}
\put(40,150){\line(1,0){8}}
\multiput(40,90)(0,-10){10}{\line(1,0){8}}

\multiput(48,150)(000,-10){16}{\line(1,0){7}} 

\put(55,96){\framebox(8,8){Z}}
\put(55,106){\framebox(8,8){Y}}
\put(55,120){\line(1,0){8}}
\put(55,130){\line(1,0){8}}
\put(59,140){\circle*{3}}
\put(55,140){\line(1,0){8}}
\put(55,150){\line(1,0){8}}
\multiput(55,90)(0,-10){10}{\line(1,0){8}}
\put(59,140){\line(0,-1){26}}
\put(59,104){\line(0,1){2}}

\multiput(63,150)(000,-10){16}{\line(1,0){7}} 

\put(70,96){\framebox(8,8){Z}}
\put(70,106){\framebox(8,8){Y}}
\put(70,120){\line(1,0){8}}
\put(74,130){\circle*{3}}
\put(70,130){\line(1,0){8}}
\put(70,136){\framebox(8,8){Z}}
\put(70,150){\line(1,0){8}}
\multiput(70,90)(0,-10){10}{\line(1,0){8}}
\put(74,136){\line(0,-1){22}}
\put(74,104){\line(0,1){2}}

\multiput(78,150)(000,-10){16}{\line(1,0){7}} 

\put(85,96){\framebox(8,8){Z}}
\put(85,106){\framebox(8,8){X}}
\put(89,120){\circle*{3}}
\put(85,120){\line(1,0){8}}
\put(85,126){\framebox(8,8){Z}}
\put(85,140){\line(1,0){8}}
\put(85,150){\line(1,0){8}}
\multiput(85,90)(0,-10){10}{\line(1,0){8}}
\put(89,126){\line(0,-1){12}}
\put(89,104){\line(0,1){2}}


\multiput(93,150)(000,-10){16}{\line(1,0){7}} 

\put(104,50){\circle*{3}}
\put(100,50){\line(1,0){8}}
\put(100,56){\framebox(8,8){X}}
\put(104,50){\line(0,1){6}}
\put(100,70){\line(1,0){8}}
\put(100,80){\line(1,0){8}}
\put(100,90){\line(1,0){8}}
\put(100,100){\line(1,0){8}}
\multiput(100,40)(0,-10){5}{\line(1,0){8}}
\multiput(100,150)(0,-10){5}{\line(1,0){8}}

\multiput(108,150)(000,-10){16}{\line(1,0){7}} 

\put(115,46){\framebox(8,8){Z}}
\put(115,56){\framebox(8,8){Y}}
\put(115,70){\line(1,0){8}}
\put(115,80){\line(1,0){8}}
\put(119,90){\circle*{3}}
\put(115,90){\line(1,0){8}}
\put(115,96){\framebox(8,8){Z}}
\multiput(115,40)(0,-10){5}{\line(1,0){8}}
\multiput(115,150)(0,-10){5}{\line(1,0){8}}
\put(119,96){\line(0,-1){32}}
\put(119,54){\line(0,1){2}}

\multiput(123,150)(000,-10){16}{\line(1,0){7}} 

\put(130,46){\framebox(8,8){Z}}
\put(130,56){\framebox(8,8){Y}}
\put(130,70){\line(1,0){8}}
\put(134,80){\circle*{3}}
\put(130,80){\line(1,0){8}}
\put(130,86){\framebox(8,8){Z}}
\put(130,100){\line(1,0){8}}
\multiput(130,40)(0,-10){5}{\line(1,0){8}}
\multiput(130,150)(0,-10){5}{\line(1,0){8}}
\put(134,86){\line(0,-1){22}}
\put(134,54){\line(0,1){2}}

\multiput(138,150)(000,-10){16}{\line(1,0){7}} 

\put(145,46){\framebox(8,8){Z}}
\put(145,56){\framebox(8,8){X}}
\put(149,70){\circle*{3}}
\put(145,70){\line(1,0){8}}
\put(145,76){\framebox(8,8){Z}}
\put(145,90){\line(1,0){8}}
\put(145,100){\line(1,0){8}}
\multiput(145,40)(0,-10){5}{\line(1,0){8}}
\multiput(145,150)(0,-10){5}{\line(1,0){8}}
\put(149,76){\line(0,-1){12}}
\put(149,54){\line(0,1){2}}


\multiput(153,150)(000,-10){16}{\line(1,0){7}} 

\put(164,0){\circle*{3}}
\put(160,0){\line(1,0){8}}
\put(160,6){\framebox(8,8){X}}
\put(164,0){\line(0,1){6}}
\put(160,20){\line(1,0){8}}
\put(160,30){\line(1,0){8}}
\put(160,40){\line(1,0){8}}
\put(160,50){\line(1,0){8}}
\multiput(160,150)(0,-10){10}{\line(1,0){8}}

\multiput(168,150)(000,-10){16}{\line(1,0){7}} 

\put(175,-4){\framebox(8,8){Z}}
\put(175,6){\framebox(8,8){Y}}
\put(175,20){\line(1,0){8}}
\put(175,30){\line(1,0){8}}
\put(179,40){\circle*{3}}
\put(175,40){\line(1,0){8}}
\put(175,46){\framebox(8,8){Z}}
\multiput(175,150)(0,-10){10}{\line(1,0){8}}
\put(179,46){\line(0,-1){32}}
\put(179,4){\line(0,1){2}}

\multiput(183,150)(000,-10){16}{\line(1,0){7}} 

\put(190,-4){\framebox(8,8){Z}}
\put(190,6){\framebox(8,8){Y}}
\put(190,20){\line(1,0){8}}
\put(194,30){\circle*{3}}
\put(190,30){\line(1,0){8}}
\put(190,36){\framebox(8,8){Z}}
\put(190,50){\line(1,0){8}}
\multiput(190,150)(0,-10){10}{\line(1,0){8}}
\put(194,36){\line(0,-1){22}}
\put(194,4){\line(0,1){2}}

\multiput(198,150)(000,-10){16}{\line(1,0){7}} 

\put(205,-4){\framebox(8,8){Z}}
\put(205,6){\framebox(8,8){X}}
\put(209,20){\circle*{3}}
\put(205,20){\line(1,0){8}}
\put(205,26){\framebox(8,8){Z}}
\put(205,40){\line(1,0){8}}
\put(205,50){\line(1,0){8}}
\multiput(205,150)(0,-10){10}{\line(1,0){8}}
\put(209,26){\line(0,-1){12}}
\put(209,4){\line(0,1){2}}

\multiput(213,150)(000,-10){16}{\line(1,0){7}} 

\end{picture}
\caption[Full encoding circuit for the 5-qubit convolutional code in
the standard polynomial form.]{Circuit for encoding the first three
qubits of a stream of quantum information with the 5-qubit
convolutional code. All the simplifications have been done. The first
physical qubit corresponds to a sacrificed logical qubit. In this
case, it could be removed since it is never involved in a quantum
gate.}\label{fig:encoding}
\end{center}
\end{figure}
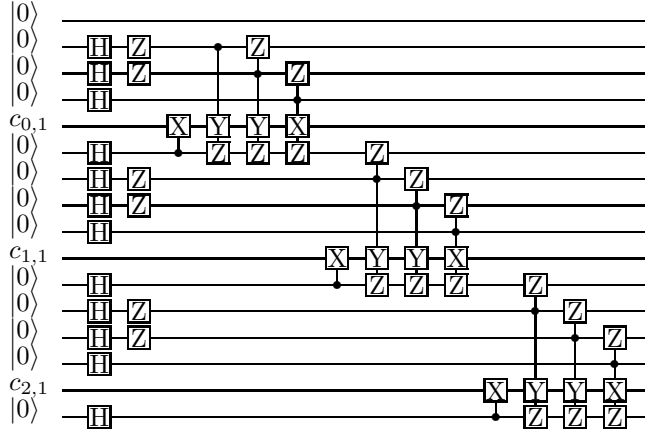

For the 5-qubit convolutional code, the full encoding circuit is
presented in Fig.~\ref{fig:encoding}, where all the simplifications
have been implemented. Here, the existence of sacrificed logical
qubits is clearly apparent: the first qubit is never involved in any
gate and does not contain any quantum information. This comes from the
finiteness of the to-be-protected sequence: at the beginning and at
the end of the stream, there are less commutation constraints for the
encoded Pauli operators imposed by the generators in $M$. Thus, it is
not surprising that there exist a finite number of encoded Pauli
operators that do not follow the convolutional structure. It is also
important to remark that there is no need to determine these operators
explicitly for deriving the encoding circuit. Setting the sacrificed
qubits to the logical $\ket 0$ state is taken care of by setting the
first $\lambda$ $n$-qubit blocks to the all-zeroes state.

\section{Error propagation and on-line decoding}\label{sec:catastrophicity}
The previous section was devoted to the derivation of the encoding
circuit for quantum convolutional codes. It showed how the standard
polynomial form for the generators of the code leads to an automated
procedure for finding an on-line encoding circuit. In this section,
the focus shifts to decoding quantum convolutional codes. The need for
a clear discussion on this issue comes from the specificity of
convolutional codes: usual decoding circuits---obtained by running the
encoding one in reverse direction---require to wait for the last
logical qubit before running them. This is not a practical option as
it would cause long transmission delays.

Here, we show that the existence of an on-line decoding circuit is
implied by a more fundamental property of the encoding operation: the
absence of catastrophic errors. These errors will be defined carefully
below, but we can already mention that they are not specific to
quantum codes. Rather they, and more generally all the error
propagation problems considered in this section, are also encountered
in the theory of classical convolutional codes~\cite{Lee97a, JZ99a}.

To build our intuition on the error propagation problems that might
arise when using convolutional codes, consider a generic encoding
circuit as derived in the previous section (see also
Fig.~\ref{fig:general_encoding}). Because of the overlap between the
generators on $m$ qubits as defined in
Eq.~(\ref{eq:general_structure}), quantum information is propagated
from one $n$-qubit block to another. As a consequence, even though the
to-be-protected stream of information is in a separable state, say
$\ket {0, \ldots, 0}$, the encoded state is not, in general, separable
with respect to any bipartite cut. In spite of their relatively simple
form---invariant by shifts of $n$-qubit---encoding circuits apply
global unitary transformations that cannot be casted in tensor
products of smaller unitary operations.

The good spreading of quantum information induced by the particular
structure of convolutional codes might in some cases have a bad
consequences: nothing prevents an error affecting a finite number of
qubits before the complete decoding of the stream to propagate
infinitely through the decoding circuit. Such error is called
catastrophic. 

\begin{definition}[Catastrophic error]
Consider an $(n,k,m)$-convolutional encoding sche\-me for protecting
$q\times k$ logical qubits. A catastrophic error is an error that
affects $O(1)$ qubits before the end of the decoding operation and
that can only be corrected by a unitary transformation whose size of
support grows with $q$, for large $q$.
\end{definition}

\begin{remark}\rm
The theory of classical convolutional codes explicitly shows the
existence of catastrophic errors for some convolutional encoders. As
these are a special case of quantum codes---their generators are tensor
products of $I$'s and $Z$'s---it proves the existence of catastrophic
errors for some quantum encoding circuits.\eend
\end{remark}

\subsection{Catastrophicity condition}
\begin{figure}[tp]
\begin{center}
\setlength{\unitlength}{0.4mm}
\begin{picture}(210, 140)

\fontsize{8pt}{9.6pt}

\multiput(0,0)(0,10){15}{\line(1,0){20}}
\multiput(30,0)(0,10){15}{\line(1,0){10}}
\multiput(50,0)(0,10){15}{\line(1,0){10}}
 \multiput(70,0)(0,10){15}{\line(1,0){10}}
\multiput(90,0)(0,10){15}{\line(1,0){10}}
\multiput(110,0)(0,10){15}{\line(1,0){10}}
\multiput(130,0)(0,10){15}{\line(1,0){10}}
\multiput(150,0)(0,10){15}{\line(1,0){10}}
\multiput(170,0)(0,10){15}{\line(1,0){10}}
\multiput(190,0)(0,10){15}{\line(1,0){20}}

\multiput(8,-2)(0,10){15}{\line(0,1){4}}
\multiput(10,8)(0,20){7}{\line(0,1){4}}

\put(20,138){\line(1,0){10}}
\put(20,138){\line(0,1){4}}
\put(30,138){\line(0,1){4}}
\multiput(20,0)(0,10){14}{\line(1,0){10}}

\put(40,118){\framebox(10,24){$U$}}
\multiput(40,0)(0,10){12}{\line(1,0){10}}

\put(60,98){\framebox(10,24){$U$}}
\multiput(60,0)(0,10){10}{\line(1,0){10}}
\multiput(60,130)(0,10){2}{\line(1,0){10}}

\put(80,78){\framebox(10,24){$U$}}
\multiput(80,0)(0,10){8}{\line(1,0){10}}
\multiput(80,110)(0,10){4}{\line(1,0){10}}

\put(100,58){\framebox(10,24){$U$}}
\multiput(100,0)(0,10){6}{\line(1,0){10}}
\multiput(100,90)(0,10){6}{\line(1,0){10}}

\put(120,38){\framebox(10,24){$U$}}
\multiput(120,0)(0,10){4}{\line(1,0){10}}
\multiput(120,70)(0,10){8}{\line(1,0){10}}

\put(140,18){\framebox(10,24){$U$}}
\multiput(140,0)(0,10){2}{\line(1,0){10}}
\multiput(140,50)(0,10){10}{\line(1,0){10}}

\put(160,-2){\framebox(10,24){$U$}}
\multiput(160,0)(0,10){0}{\line(1,0){10}}
\multiput(160,30)(0,10){12}{\line(1,0){10}}

\put(180,2){\line(1,0){10}}
\put(180,2){\line(0,-1){4}}
\put(190,2){\line(0,-1){4}}
\multiput(180,10)(0,10){14}{\line(1,0){10}}

\end{picture}
\caption[Typical encoding circuit for a convolutional code.]  {Typical
encoding circuit for a convolutional code. The circuit is run from
left to right. Horizontal lines of a given type (i.e.\ with single or
double vertical bar) always represent the same number of qubits. The
unitary operation $U$ is implemented as a series of elementary gates
acting only on the qubits with which it intersects.}
\label{fig:general_encoding}
\end{center}
\end{figure}
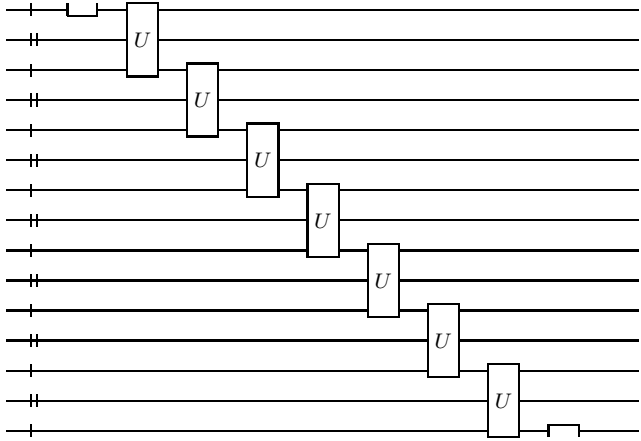

In this paragraph, we will find a catastrophicity condition for
convolutional encoders without relying on the stabilizer description
of the code. Instead, we simply assume a generic form for the encoding
operation of $q\times k$ to-be-protected qubits:
\begin{equation}
C(q) = T_{\mathrm{erm}} \times D^{q-1}[U] \times \ldots \times D[U] \times U
\times I_{\mathrm{nit}}, \label{eq:encoding_circuit}
\end{equation}
where $I_{\mathrm{nit}}$ and $T_{\mathrm{erm}}$ are two fixed unitary
transformations, respectively the initia\-li\-zation---acting at the
beginning of the to-be-protected stream of information---, and the
termination---acting on the last qubits of the stream.\footnote{The
delay operator, initially introduced only for elements of the Pauli
group with finite support, is easily generalized to handle unitary
matrices with finite support.}  The unitary $U$ has a finite support
independent of $q$. In the standard encoding presented in the previous
section, $U$ corresponds to the encoding of $k$ consecutive qubits
containing information---i.e.\ it corresponds to applying some
$\overline X$'s and some $M_{j,i}$'s. The presence of
$I_{\mathrm{nit}}$ and $T_{\mathrm{erm}}$ is due to the sacrificed
logical qubits at the beginning and at the end of the encoded
stream. The typical arrangement of the unitary operations $D^{i}[U]$
far from the beginning and the end of the stream of information is
depicted in Fig.~\ref{fig:general_encoding}.


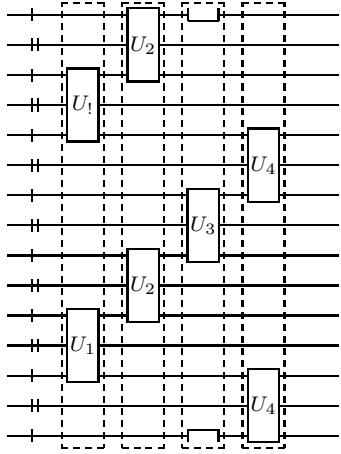
\begin{figure}[tp]
\begin{center}

\setlength{\unitlength}{0.4mm}

\begin{picture}(110, 140)

\fontsize{8pt}{9.6pt}

\multiput(0,0)(0,10){15}{\line(1,0){20}}
\multiput(30,0)(0,10){15}{\line(1,0){10}}
\multiput(50,0)(0,10){15}{\line(1,0){10}}
\multiput(70,0)(0,10){15}{\line(1,0){10}}
\multiput(90,0)(0,10){15}{\line(1,0){20}}

\multiput(8,-2)(0,10){15}{\line(0,1){4}}
\multiput(10,8)(0,20){7}{\line(0,1){4}}

\multiput(8,-2)(0,10){15}{\line(0,1){4}}
\multiput(10,8)(0,20){7}{\line(0,1){4}}

\put(20,18){\framebox(10,24){$U_1$}}
\put(20,98){\framebox(10,24){$U_!$}}
\multiput(20,0)(0,10){2}{\line(1,0){10}}
\multiput(20,50)(0,10){5}{\line(1,0){10}}
\multiput(20,130)(0,10){2}{\line(1,0){10}}
\put(18,-4){\dashbox{2}(14,148){}}

\put(40,38){\framebox(10,24){$U_2$}}
\put(40,118){\framebox(10,24){$U_2$}}
\multiput(40,00)(0,10){4}{\line(1,0){10}}
\multiput(40,70)(0,10){5}{\line(1,0){10}}
\put(38,-4){\dashbox{2}(14,148){}}

\put(60,2){\line(1,0){10}}
\put(60,2){\line(0,-1){4}}
\put(70,2){\line(0,-1){4}}
\put(60,58){\framebox(10,24){$U_3$}}
\put(60,138){\line(1,0){10}}
\put(60,138){\line(0,1){4}}
\put(70,138){\line(0,1){4}}
\multiput(60,10)(0,10){5}{\line(1,0){10}}
\multiput(60,90)(0,10){5}{\line(1,0){10}}
\put(58,-4){\dashbox{2}(14,148){}}

\put(80,-2){\framebox(10,24){$U_4$}}
\put(80,78){\framebox(10,24){$U_4$}}
\multiput(80,30)(0,10){5}{\line(1,0){10}}
\multiput(80,110)(0,10){4}{\line(1,0){10}}
\put(78,-4){\dashbox{2}(14,148){}}

\end{picture}
\caption[Example of pearl-necklace structure for the encoding
circuit.]  {Example of pearl-necklace structure for the encoding
circuit. We have depicted four layers of unitaries, $U_1$ through
$U_4$. Here, the condition of commutation inside a layer is guaranteed
by the disjointness of the support of the different unitaries
$\{D^j[U_i]\}_j$.}
\label{fig:pearl_necklace}
\end{center}
\end{figure}

\begin{proposition}
A quantum convolutional encoder is non-catastrophic if and only if the
encoding operation $C(q)$ can be decomposed in the following way for
large $q$:
\begin{equation}
C(q) = \tilde T_{\mathrm{erm}}(q)\times 
\left (\prod_{i = 0}^{\lfloor q/l_t \rfloor} D^{i l_t}[U_t] \right) \times 
\ldots \times 
\left (\prod_{i = 0}^{\lfloor q/l_1 \rfloor} D^{i l_1}[U_1] \right) \times 
\tilde I_{\mathrm{nit}}(q),\label{eq:non_catastrophic}
\end{equation}
where $\tilde I_{\mathrm{nit}}(q)$ and $\tilde T_{\mathrm{erm}}(q)$
are modified initialization and termination steps which can vary with
$q$, but whose support is bounded; $\{U_j\}_j$ is a finite set of
unitary operators independent of $q$---thus with bounded
support---such that $D^i[U_j]$ and $D^{i'}[U_j]$ commute; and $l_j$'s
are integers independent of $q$.
\end{proposition}
Even though this condition might seem at first sight quite
complicated, it corresponds to a reordering of the unitaries---or
gates---in the quantum circuit which is easy to understand. The new
circuit must have the following form: first an initialization step is
performed;\footnote{Because of possible side effects $\tilde
I_{\mathrm{nit}}(q)$ can depend on $q$ but the size of its support
must be of order 1 and it can act non-trivially only on the first few
qubits} then, there are $t$ layers of unitaries (each of them made out
of a single unitary, e.g.\ $U_i$, and its $n$-qubit shifted versions)
such that the gates inside a layer commute with each other; finally it
is followed by a termination step, $\tilde T_{\mathrm{erm}}(q)$ with
bounded support. This structure resembles a pearl-necklace as it can
be seen on Fig.~\ref{fig:pearl_necklace}.


\begin{proof}[Sufficiency]\rm
To simplify the discussion, we will consider the case where the error
$E$ occurs before the beginning of the decoding operation. This is not
general, since the definition of non-catastrophicity also imposes to
consider errors occurring on a partially decoded stream. Nonetheless,
the proof presented here can be easily adapted for this other case.

Here, we have to show that for $q$ large, whenever $E$ has bounded
support, $C(q)^\+ E C(q)$ has a bounded support as well. Since $\tilde
T_{\mathrm{erm}}(q)$ has a bounded support at the end of the stream,
it is always possible to increase $q$ such that $E$ and $\tilde
T_{\mathrm{erm}}(q)$ commute. Therefore, after simplifying $C(q)^\+ E
C(q)$ by $\tilde T_{\mathrm{erm}}$, we have:
\begin{eqnarray}
C(q)^\+ E C(q) & = & 
\tilde I_{\mathrm{nit}}(q)^\+ \times \left (\prod_{i = 0}^{\lfloor q/l_1 \rfloor}
D^{i l_1}[U_1^\+] \right) \times \ldots \times \left (\prod_{i =
0}^{\lfloor q/l_t \rfloor} D^{i l_t}[U_t^\+] \right) \times E \times \nonumber\\ 
&& \times \left (\prod_{i = 0}^{\lfloor q/l_t \rfloor} D^{i l_t}[U_t]
\right) \times \ldots \times \left (\prod_{i = 0}^{\lfloor q/l_1
\rfloor} D^{i l_1}[U_1] \right) \times \tilde I_{\mathrm{nit}}(q). 
\end{eqnarray}
Similarly, in the above equation all the $D^{il_t}[U_t]$ whose support
does not intersect the one of $E$ commute with it and can be
simplified (recall also that the $D^{il_t}[U_t]$ also commute with
each other). Only a finite number of the $D^{il_t}[U_t]$'s remain, say
$\{D^{il_t}[U_t]\}_{i \in I_t}$. Note that for $q$ large, this number
is independent of $q$. We thus have
\begin{eqnarray}
C(q)^\+ E C(q) & = & 
\tilde I_{\mathrm{nit}}(q)^\+ \times \left (\prod_{i = 0}^{\lfloor q/l_1 \rfloor}
D^{i l_1}[U_1^\+] \right) \times \ldots \times \left (\prod_{i =
0}^{\lfloor q/l_{t-1} \rfloor} D^{i l_{t-1}}[U_{t-1}^\+] \right) \times \nonumber \\
&& \times E_1 \times \\ 
&& \times \left (\prod_{i = 0}^{\lfloor q/l_{t-1} \rfloor} D^{i l_{t-1}}[U_{t-1}]
\right) \times \ldots \times \left (\prod_{i = 0}^{\lfloor q/l_1
\rfloor} D^{i l_1}[U_1] \right) \times \tilde I_{\mathrm{nit}}(q) \nonumber
\end{eqnarray}
where $E_1 = \left(\prod_{i \in I_t} D^{i l_t}[U_t^\+]\right) \times E
\times \left(\prod_{i \in I_t} D^{i l}[U_t]\right)$ has a bounded
support, independent of $q$. The rest of the proof follows immediately
by applying the same technique to the remaining layers: another step
generates $E_2$, by considering $E_1$ instead of $E$ and $U_{t-1}$
instead of $U_t$. Following the same arguments, $E_2$ has a bounded
support independent of $q$ and so will $E_3, \ldots, E_t$. Thus it
proves that $C(q)^\+ E C(q) = \tilde I_{\mathrm{nit}}(q)^\+ E_t \tilde
I_{\mathrm{nit}}(q)$ has bounded support.\eend
\end{proof}

\begin{proof}[Necessity]\rm
To prove that this condition is necessary, we will show that a
non-catastrophic encoding operation $C(q)$ can be put in the special
form of Eq.~(\ref{eq:non_catastrophic}), for $q$ large. The outline of
the proof is the following: we will work on the circuit of the
decoding operation $C(q)^\+$, obtained by running the encoding circuit
in the reverse direction (see Fig.~\ref{fig:general_decoding}). Our
goal is to convert this decoding circuit into an equivalent one which
displays the pearl-necklace structure. To do so, we will consider a
possible---but yet very particular---error which could occur on the
physical qubits during the transmission. The chosen error indeed
corresponds to a local reordering of the unitaries in $C(q)^\+$. Since
the encoding is supposed to have no catastrophic errors, this local
reordering can be compensated by applying a unitary operation with
finite support after complete decoding. This will give us an identity
between two decoding circuits, which we can apply as many times as
required to arrive at the pearl-necklace structure.

\begin{figure}[tp]
\begin{center}

\setlength{\unitlength}{0.4mm}

\begin{picture}(210, 140)

\fontsize{8pt}{9.6pt}

\multiput(0,0)(0,10){15}{\line(1,0){20}}
\multiput(30,0)(0,10){15}{\line(1,0){10}}
\multiput(50,0)(0,10){15}{\line(1,0){10}}
\multiput(70,0)(0,10){15}{\line(1,0){10}}
\multiput(90,0)(0,10){15}{\line(1,0){10}}
\multiput(110,0)(0,10){15}{\line(1,0){10}}
\multiput(130,0)(0,10){15}{\line(1,0){10}}
\multiput(150,0)(0,10){15}{\line(1,0){10}}
\multiput(170,0)(0,10){15}{\line(1,0){10}}
\multiput(190,0)(0,10){15}{\line(1,0){20}}

\multiput(8,-2)(0,10){15}{\line(0,1){4}}
\multiput(10,8)(0,20){7}{\line(0,1){4}}

\put(20,2){\line(1,0){10}}
\put(20,2){\line(0,-1){4}}
\put(30,2){\line(0,-1){4}}
\multiput(20,10)(0,10){14}{\line(1,0){10}}

\put(40,-2){\framebox(10,24){$U^\+$}}
\multiput(40,0)(0,10){0}{\line(1,0){10}}
\multiput(40,30)(0,10){12}{\line(1,0){10}}

\put(60,18){\framebox(10,24){$U^\+$}}
\multiput(60,0)(0,10){2}{\line(1,0){10}}
\multiput(60,50)(0,10){10}{\line(1,0){10}}

\put(80,38){\framebox(10,24){$U^\+$}}
\multiput(80,0)(0,10){4}{\line(1,0){10}}
\multiput(80,70)(0,10){8}{\line(1,0){10}}

\put(100,58){\framebox(10,24){$U^\+$}}
\multiput(100,0)(0,10){6}{\line(1,0){10}}
\multiput(100,90)(0,10){6}{\line(1,0){10}}

\put(120,78){\framebox(10,24){$U^\+$}}
\multiput(120,0)(0,10){8}{\line(1,0){10}}
\multiput(120,110)(0,10){4}{\line(1,0){10}}

\put(140,98){\framebox(10,24){$U^\+$}}
\multiput(140,0)(0,10){10}{\line(1,0){10}}
\multiput(140,130)(0,10){2}{\line(1,0){10}}

\put(160,118){\framebox(10,24){$U^\+$}}
\multiput(160,0)(0,10){12}{\line(1,0){10}}

\put(180,138){\line(1,0){10}}
\put(180,138){\line(0,1){4}}
\put(190,138){\line(0,1){4}}
\multiput(180,0)(0,10){14}{\line(1,0){10}}

\end{picture}
\caption[Typical decoding circuit for a convolutional code.]  {Typical
decoding circuit for a convolutional code. The circuit is obtained by
running the encoding circuit in reverse direction and with appropriate
Hermitian conjugates.}
\label{fig:general_decoding}
\end{center}
\end{figure}
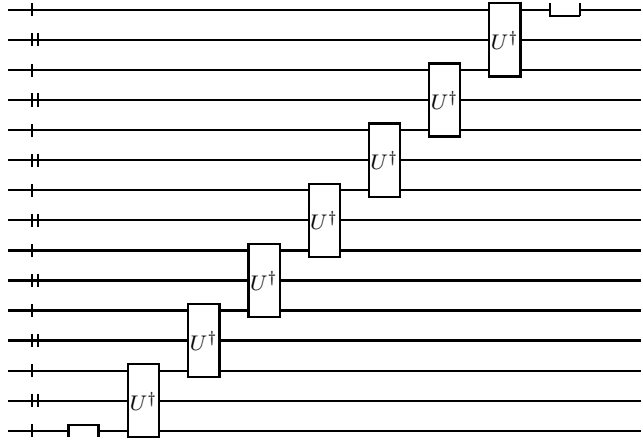

More specifically, consider the decoding unitary operation,
\begin{equation}
C(q)^\+ = I_{\mathrm{nit}}^\+ \times U^\+ \times D[U^\+] \times \ldots \times
D^{q-1}[U^\+] \times T_{\mathrm{erm}}^\+.\label{eq:typical_decoding}
\end{equation}
We define the integer $l$ such that $U$ and $D^i[U]$ have disjoint
support for $|i| > l$.\footnote{This integer exists because $U$ has
finite support.}  The circuit identity that will be derived is:
\begin{equation}
C(q)^\+ = D^{q-l'}[V^\+] \times \tilde C(q)^\+, 
\end{equation}
where $V$ has finite support extending on $l'$ $n$-qubit blocks, and
where $\tilde C(q)^\+$ is obtained from $C(q)^\+$ by locally
reordering its last $2l+1$ unitaries $U$:
\begin{eqnarray}
\tilde C(q)^\+ & = &I_{\mathrm{nit}}^\+ \times U^\+ \times \ldots \times
D^{q-2l-3}[U^\+] \times \nonumber \\ 
&& \times D^{q-2l-2}[D^{l+1}[U^\+] \times U^\+] \times D^{q-2l-1}[D^{l+1}[U^\+] \times U^\+] \times \ldots \\
&& \ldots \times D^{q-l-2}(D^{l+1}[U^\+] \times U^\+) \times T_{\mathrm{erm}}^\+.\nonumber
\end{eqnarray}

Consider $E$, a unitary operation, defined by:
\begin{eqnarray}
E & = & (D^{l+1}[U^\+] \times U^\+) \times D[D^{l+1}[U^\+] \times
U^\+] \times \ldots \times D^{l}[D^{l+1}[U^\+] \times U^\+] \times
\nonumber \\ && \times D^{2l+1}[U] \times D^{2l}[U] \times \ldots
\times D[U] \times U.\label{eq:error_commutator}
\end{eqnarray}
An illustration of the arrangement of the unitaries in $E$ is
presented on Fig.~\ref{fig:local_error} for $l = 1$.  By construction,
$E$ satisfies:
\begin{eqnarray}
\tilde C(q)^\+ & = & I_{\mathrm{nit}}^\+ \times U^\+ \times \ldots
\times D^{q-2l-3}[U^\+] \times \ldots \nonumber \\ && \times
D^{q-2l-2}[E] \times \label{eq:reordered}\\ && \times D^{q-2l-2}[U^\+] \times
D^{q-2l-1}[U^\+] \times \ldots \times D^{q-1}[U^\+] \times
T_{\mathrm{erm}}^\+, \nonumber
\end{eqnarray}
which simply corresponds to the initial decoding operation $C(q)^\+$
with an error $E$ happening between the unitaries $D^{q-2l-3}[U]$ and
$D^{q-2l-2}[U]$. Since, the encoding is non-catastrophic, there exists
a unitary $V^\+$ with finite support---also obviously independent of
$q$---such that $C(q)^\+ = D^{q-l'}[V^\+] \times \tilde C(q)$, where
$l'$ is the size of the support of $V$ counted in number of $n$-qubit
blocks, which gives the circuit identity (see
Figs.~\ref{fig:local_reordering1}~\&~\ref{fig:local_reordering2} for
the local reordering implied by
Eqs.~(\ref{eq:typical_decoding}--\ref{eq:reordered}).

\begin{figure}[tp]
\begin{center}

\setlength{\unitlength}{0.4mm}

\begin{picture}(150, 90)

\fontsize{8pt}{9.6pt}

\multiput(0,0)(0,10){9}{\line(1,0){20}}
\multiput(30,0)(0,10){9}{\line(1,0){10}}
\multiput(50,0)(0,10){9}{\line(1,0){10}}
\multiput(70,0)(0,10){9}{\line(1,0){10}}
\multiput(90,0)(0,10){9}{\line(1,0){10}}
\multiput(110,0)(0,10){9}{\line(1,0){10}}
\multiput(130,0)(0,10){9}{\line(1,0){20}}

\multiput(8,-2)(0,10){9}{\line(0,1){4}}
\multiput(10,8)(0,20){4}{\line(0,1){4}}

\put(20,58){\framebox(10,24){$U$}}
\multiput(20,90)(0,10){0}{\line(1,0){10}}
\multiput(20,00)(0,10){6}{\line(1,0){10}}

\put(40,38){\framebox(10,24){$U$}}
\multiput(40,70)(0,10){2}{\line(1,0){10}}
\multiput(40,0)(0,10){4}{\line(1,0){10}}

\put(60,18){\framebox(10,24){$U$}}
\multiput(60,50)(0,10){4}{\line(1,0){10}}
\multiput(60,0)(0,10){2}{\line(1,0){10}}

\put(80,-2){\framebox(10,24){$U$}}
\multiput(80,30)(0,10){6}{\line(1,0){10}}
\multiput(80,0)(0,10){0}{\line(1,0){10}}

\put(100,-2){\framebox(10,24){$U^\+$}}
\multiput(100,30)(0,10){1}{\line(1,0){10}}

\put(100,38){\framebox(10,24){$U^\+$}}
\multiput(100,70)(0,10){2}{\line(1,0){10}}

\put(120,18){\framebox(10,24){$U^\+$}}
\multiput(120,0)(0,10){2}{\line(1,0){10}}

\put(120,58){\framebox(10,24){$U^\+$}}
\multiput(120,50)(0,10){1}{\line(1,0){10}}

\end{picture}
\caption[Error operation $E$ as defined in
Eq.~(\ref{eq:error_commutator})] {Error operation $E$ as defined in
Eq.~(\ref{eq:error_commutator}). Here, $l = 1$ because $D^i[U]$
commutes with $U$ for $i > 1$. When introduced in the decoding
circuit, such operation induces a local reordering of the unitaries
$U^\+$.}
\label{fig:local_error}
\end{center}
\end{figure}
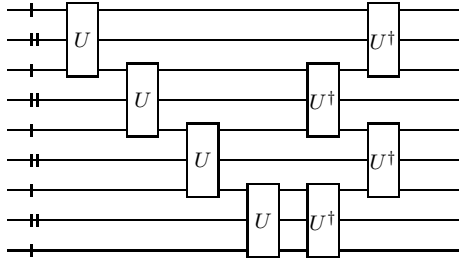

\begin{figure}[tp]
\begin{center}

\setlength{\unitlength}{0.4mm}

\begin{picture}(250, 140)

\fontsize{8pt}{9.6pt}

\multiput(0,0)(0,10){15}{\line(1,0){20}}
\multiput(30,0)(0,10){15}{\line(1,0){10}}
\multiput(50,0)(0,10){15}{\line(1,0){10}}
\multiput(70,0)(0,10){15}{\line(1,0){10}}
\multiput(90,0)(0,10){15}{\line(1,0){10}}
\multiput(110,0)(0,10){15}{\line(1,0){10}}
\multiput(130,0)(0,10){15}{\line(1,0){10}}
\multiput(150,0)(0,10){15}{\line(1,0){10}}
\multiput(170,0)(0,10){15}{\line(1,0){10}}
\multiput(190,0)(0,10){15}{\line(1,0){10}}
\multiput(210,0)(0,10){15}{\line(1,0){10}}
\multiput(230,0)(0,10){15}{\line(1,0){20}}

\multiput(8,-2)(0,10){15}{\line(0,1){4}}
\multiput(10,8)(0,20){7}{\line(0,1){4}}

\put(20,2){\line(1,0){10}}
\put(20,2){\line(0,-1){4}}
\put(30,2){\line(0,-1){4}}
\multiput(20,10)(0,10){14}{\line(1,0){10}}

\put(40,-2){\framebox(10,24){$U^\+$}}
\multiput(40,0)(0,10){0}{\line(1,0){10}}
\multiput(40,30)(0,10){12}{\line(1,0){10}}

\put(60,18){\framebox(10,24){$U^\+$}}
\multiput(60,0)(0,10){2}{\line(1,0){10}}
\multiput(60,50)(0,10){10}{\line(1,0){10}}

\put(80, -2){\framebox(10,44){$E$}}
\multiput(80,0)(0,10){0}{\line(1,0){10}}
\multiput(80,50)(0,10){10}{\line(1,0){10}}

\put(100,38){\framebox(10,24){$U^\+$}}
\multiput(100,0)(0,10){4}{\line(1,0){10}}
\multiput(100,70)(0,10){8}{\line(1,0){10}}

\put(120,58){\framebox(10,24){$U^\+$}}
\multiput(120,0)(0,10){6}{\line(1,0){10}}
\multiput(120,90)(0,10){6}{\line(1,0){10}}

\put(140,78){\framebox(10,24){$U^\+$}}
\multiput(140,0)(0,10){8}{\line(1,0){10}}
\multiput(140,110)(0,10){4}{\line(1,0){10}}

\put(160,98){\framebox(10,24){$U^\+$}}
\multiput(160,0)(0,10){10}{\line(1,0){10}}
\multiput(160,130)(0,10){2}{\line(1,0){10}}

\put(180,118){\framebox(10,24){$U^\+$}}
\multiput(180,0)(0,10){12}{\line(1,0){10}}

\put(200,138){\line(1,0){10}}
\put(200,138){\line(0,1){4}}
\put(210,138){\line(0,1){4}}
\multiput(200,0)(0,10){14}{\line(1,0){10}}

\put(220,-2){\framebox(10,74){$V$}}
\multiput(220,0)(0,10){0}{\line(1,0){10}}
\multiput(220,80)(0,10){7}{\line(1,0){10}}

\end{picture}
\caption[Derivation of a circuit identity for decoding.]  {Derivation
of a circuit identity for decoding. Because there is no catastrophic
error, the effect of applying $E$ as defined in
Eq.~(\ref{fig:local_error}) in the decoding circuit can be corrected
by a unitary operation $V$ with finite support: this circuit induces
the same unitary transformation on the received stream of
information.}
\label{fig:local_reordering1}
\end{center}
\end{figure}
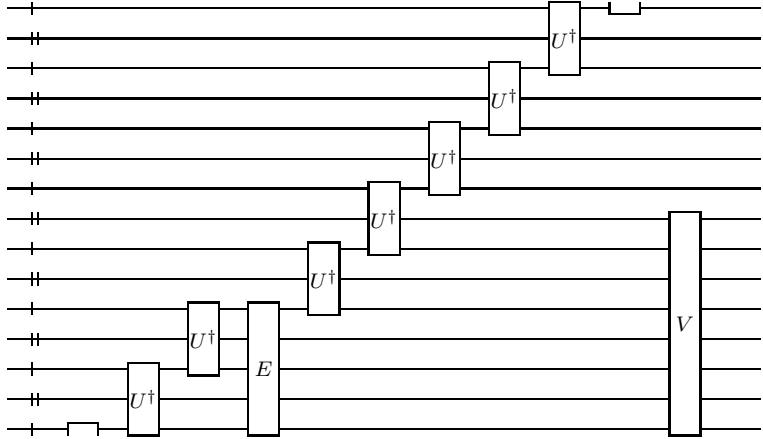

\begin{figure}[tp]
\begin{center}

\setlength{\unitlength}{0.4mm}

\begin{picture}(190, 140)

\fontsize{8pt}{9.6pt}

\multiput(0,0)(0,10){15}{\line(1,0){20}}
\multiput(30,0)(0,10){15}{\line(1,0){10}}
\multiput(50,0)(0,10){15}{\line(1,0){10}}
\multiput(70,0)(0,10){15}{\line(1,0){10}}
\multiput(90,0)(0,10){15}{\line(1,0){10}}
\multiput(110,0)(0,10){15}{\line(1,0){10}}
\multiput(130,0)(0,10){15}{\line(1,0){10}}
\multiput(150,0)(0,10){15}{\line(1,0){10}}
\multiput(170,0)(0,10){15}{\line(1,0){20}}

\multiput(8,-2)(0,10){15}{\line(0,1){4}}
\multiput(10,8)(0,20){7}{\line(0,1){4}}

\put(20,18){\framebox(10,24){$U^\+$}}
\multiput(20,0)(0,10){2}{\line(1,0){10}}
\multiput(20,50)(0,10){10}{\line(1,0){10}}

\put(40,38){\framebox(10,24){$U^\+$}}
\multiput(40,00)(0,10){4}{\line(1,0){10}}
\multiput(40,70)(0,10){8}{\line(1,0){10}}

\put(60,2){\line(1,0){10}}
\put(60,2){\line(0,-1){4}}
\put(70,2){\line(0,-1){4}}
\put(60,58){\framebox(10,24){$U^\+$}}
\multiput(60,10)(0,10){5}{\line(1,0){10}}
\multiput(60,90)(0,10){6}{\line(1,0){10}}

\put(80,-2){\framebox(10,24){$U^\+$}}
\put(80,78){\framebox(10,24){$U^\+$}}
\multiput(80,30)(0,10){5}{\line(1,0){10}}
\multiput(80,110)(0,10){4}{\line(1,0){10}}

\put(100,98){\framebox(10,24){$U^\+$}}
\multiput(100,0)(0,10){10}{\line(1,0){10}}
\multiput(100,130)(0,10){2}{\line(1,0){10}}

\put(120,118){\framebox(10,24){$U^\+$}}
\multiput(120,0)(0,10){12}{\line(1,0){10}}

\put(140,138){\line(1,0){10}}
\put(140,138){\line(0,1){4}}
\put(150,138){\line(0,1){4}}
\multiput(140,0)(0,10){14}{\line(1,0){10}}

\put(160,-2){\framebox(10,74){$V$}}
\multiput(160,0)(0,10){0}{\line(1,0){10}}
\multiput(160,80)(0,10){7}{\line(1,0){10}}

\end{picture}
\caption[Local reordering in the decoding circuit.]  {Local reordering
in the decoding circuit. By using the specific form of $E$, this
circuit is equivalent to the ones given in
Figs.~\ref{fig:general_decoding}~\&~\ref{fig:local_reordering2}}
\label{fig:local_reordering2}
\end{center}
\end{figure}
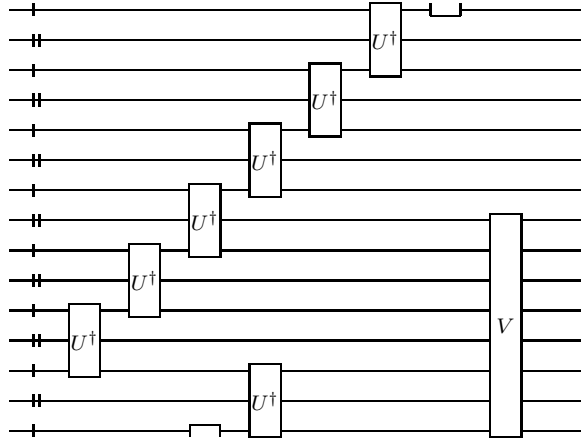

Moreover, this identity concerns only the unitary operations around
the position where $E$ is applied.  It is then possible to apply it at
repeated intervals---e.g.\ separated from $\max(l,l')+1$ $n$-qubit
blocks---in the decoding circuit.  It is then straightforward to show
that $\tilde C(q)^\+$---and similarly $\tilde C(q)$---has the form of
Eq.~(\ref{eq:non_catastrophic}), and to conclude the proof (see
Figs.~\ref{fig:global_reordering1}~\&~\ref{fig:global_reordering2}).\eend

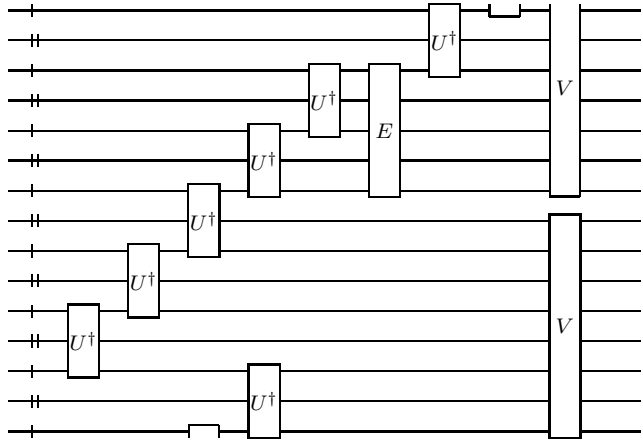
\begin{figure}[tp]
\begin{center}

\setlength{\unitlength}{0.4mm}

\begin{picture}(210, 140)

\fontsize{8pt}{9.6pt}

\multiput(0,0)(0,10){15}{\line(1,0){20}}
\multiput(30,0)(0,10){15}{\line(1,0){10}}
\multiput(50,0)(0,10){15}{\line(1,0){10}}
\multiput(70,0)(0,10){15}{\line(1,0){10}}
\multiput(90,0)(0,10){15}{\line(1,0){10}}
\multiput(110,0)(0,10){15}{\line(1,0){10}}
\multiput(130,0)(0,10){15}{\line(1,0){10}}
\multiput(150,0)(0,10){15}{\line(1,0){10}}
\multiput(170,0)(0,10){15}{\line(1,0){10}}
\multiput(190,0)(0,10){15}{\line(1,0){20}}

\multiput(8,-2)(0,10){15}{\line(0,1){4}}
\multiput(10,8)(0,20){7}{\line(0,1){4}}

\multiput(8,-2)(0,10){15}{\line(0,1){4}}
\multiput(10,8)(0,20){7}{\line(0,1){4}}

\put(20,18){\framebox(10,24){$U^\+$}}
\multiput(20,0)(0,10){2}{\line(1,0){10}}
\multiput(20,50)(0,10){10}{\line(1,0){10}}

\put(40,38){\framebox(10,24){$U^\+$}}
\multiput(40,00)(0,10){4}{\line(1,0){10}}
\multiput(40,70)(0,10){8}{\line(1,0){10}}

\put(60,2){\line(1,0){10}}
\put(60,2){\line(0,-1){4}}
\put(70,2){\line(0,-1){4}}
\put(60,58){\framebox(10,24){$U^\+$}}
\multiput(60,10)(0,10){5}{\line(1,0){10}}
\multiput(60,90)(0,10){6}{\line(1,0){10}}

\put(80,-2){\framebox(10,24){$U^\+$}}
\put(80,78){\framebox(10,24){$U^\+$}}
\multiput(80,30)(0,10){5}{\line(1,0){10}}
\multiput(80,110)(0,10){4}{\line(1,0){10}}

\put(100,98){\framebox(10,24){$U^\+$}}
\multiput(100,0)(0,10){10}{\line(1,0){10}}
\multiput(100,130)(0,10){2}{\line(1,0){10}}

\put(120, 78){\framebox(10,44){$E$}}
\multiput(120,0)(0,10){8}{\line(1,0){10}}
\multiput(120, 130)(0,10){2}{\line(1,0){10}}

\put(140,118){\framebox(10,24){$U^\+$}}
\multiput(140,0)(0,10){12}{\line(1,0){10}}

\put(160,138){\line(1,0){10}}
\put(160,138){\line(0,1){4}}
\put(170,138){\line(0,1){4}}
\multiput(160,0)(0,10){14}{\line(1,0){10}}

\put(180,-2){\framebox(10,74){$V$}}
\put(180,78){\line(1,0){10}}
\put(180,78){\line(0,1){64}}
\put(190,78){\line(0,1){64}}
\put(182,112){\makebox{$V$}}

\end{picture}
\caption[Global reordering of the decoding circuit.]  {Global
reordering of the decoding circuit. Exploiting the circuit identity
described in Fig.~\ref{fig:local_reordering2}, the fact that it
corresponds to a local reordering only (i.e.\ only a finite number of
unitaries with bounded support are involved in this identity), and the
invariance of the initial decoding circuit by $n$-qubit shifts, it is
possible to induce local reorderings at regular intervals in the
decoding circuit.}
\label{fig:global_reordering1}
\end{center}
\end{figure}

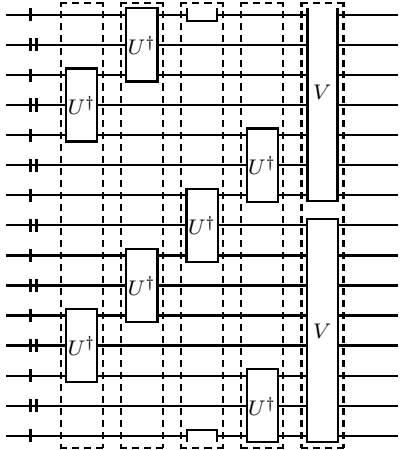
\begin{figure}[tp]
\begin{center}

\setlength{\unitlength}{0.4mm}

\begin{picture}(130, 140)

\fontsize{8pt}{9.6pt}

\multiput(0,0)(0,10){15}{\line(1,0){20}}
\multiput(30,0)(0,10){15}{\line(1,0){10}}
\multiput(50,0)(0,10){15}{\line(1,0){10}}
\multiput(70,0)(0,10){15}{\line(1,0){10}}
\multiput(90,0)(0,10){15}{\line(1,0){10}}
\multiput(110,0)(0,10){15}{\line(1,0){20}}

\multiput(8,-2)(0,10){15}{\line(0,1){4}}
\multiput(10,8)(0,20){7}{\line(0,1){4}}

\multiput(8,-2)(0,10){15}{\line(0,1){4}}
\multiput(10,8)(0,20){7}{\line(0,1){4}}

\put(20,18){\framebox(10,24){$U^\+$}}
\put(20,98){\framebox(10,24){$U^\+$}}
\multiput(20,0)(0,10){2}{\line(1,0){10}}
\multiput(20,50)(0,10){5}{\line(1,0){10}}
\multiput(20,130)(0,10){2}{\line(1,0){10}}
\put(18,-4){\dashbox{2}(14,148){}}

\put(40,38){\framebox(10,24){$U^\+$}}
\put(40,118){\framebox(10,24){$U^\+$}}
\multiput(40,00)(0,10){4}{\line(1,0){10}}
\multiput(40,70)(0,10){5}{\line(1,0){10}}
\put(38,-4){\dashbox{2}(14,148){}}

\put(60,2){\line(1,0){10}}
\put(60,2){\line(0,-1){4}}
\put(70,2){\line(0,-1){4}}
\put(60,58){\framebox(10,24){$U^\+$}}
\put(60,138){\line(1,0){10}}
\put(60,138){\line(0,1){4}}
\put(70,138){\line(0,1){4}}
\multiput(60,10)(0,10){5}{\line(1,0){10}}
\multiput(60,90)(0,10){5}{\line(1,0){10}}
\put(58,-4){\dashbox{2}(14,148){}}

\put(80,-2){\framebox(10,24){$U^\+$}}
\put(80,78){\framebox(10,24){$U^\+$}}
\multiput(80,30)(0,10){5}{\line(1,0){10}}
\multiput(80,110)(0,10){4}{\line(1,0){10}}
\put(78,-4){\dashbox{2}(14,148){}}

\put(100,-2){\framebox(10,74){$V$}}
\put(100,78){\line(1,0){10}}
\put(100,78){\line(0,1){64}}
\put(110,78){\line(0,1){64}}
\put(102,112){\makebox{$V$}}
\put(98,-4){\dashbox{2}(14,148){}}

\end{picture}
\caption[Pearl-necklace structure after global reordering of the
decoding circuit.]  {Pearl-necklace structure after global reordering
of the decoding circuit. Each layer of the structure is identified by
a dashed box. The necessity of introducing new definitions for the
initialization and termination steps, $I_{\mathrm{nit}}$ and
$T_{\mathrm{erm}}$, is due to the impossibility of applying the local
reordering when few $U^\+$'s remain at the beginning or at the end of
the decoding circuit (less than the number of $n$-qubit blocks
involved in the support of $V$).}  \label{fig:global_reordering2} 
\end{center}
\end{figure}
\end{proof}



\begin{remark}\rm
Note also, that this demonstrates the possibility of on-line decoding
for non-catastrophic quantum convolutional codes: in this form, the
``directionality'' of the quantum circuit which imposed to begin the
decoding at the end of the received stream disappeared.\eend
\end{remark}

\begin{figure}[tp]
\begin{center}
\begin{picture}(125,170)

\multiput(-20,150)(0,-10){4}{\makebox{$\ket 0$}}
\multiput(-20,100)(0,-10){4}{\makebox{$\ket 0$}}
\multiput(-20, 50)(0,-10){4}{\makebox{$\ket 0$}}
\put(-20,0){\makebox{$\ket 0$}}

\put(-20,110){\makebox{$c_{0,1}$}}
\put(-20, 60){\makebox{$c_{1,1}$}}
\put(-20, 10){\makebox{$c_{2,1}$}}

\multiput(0,150)(000,-10){16}{\line(1,0){10}} 

\multiput(10,136)(000,-10){3}{\framebox(8,8){H}}
\multiput(10, 96)(000,-10){4}{\framebox(8,8){H}}
\multiput(10, 46)(000,-10){4}{\framebox(8,8){H}}
\multiput(10, 110)(000,-50){3}{\line(1,0){8}}
\put(10,150){\line(1,0){10}}
\put(10,-4){\framebox(8,8){H}}

\put(8,-6){\dashbox(12,112){}}

\multiput(18,150)(000,-10){16}{\line(1,0){7}} 

\multiput(25, 150)(000,-50){4}{\line(1,0){8}}
\multiput(25, 136)(000,-50){3}{\framebox(8,8){Z}}
\multiput(25, 126)(000,-50){3}{\framebox(8,8){Z}}
\multiput(25, 120)(000,-50){3}{\line(1,0){8}}
\multiput(25, 110)(000,-50){3}{\line(1,0){8}}

\put(23,-6){\dashbox(12,152){}}

\multiput(33,150)(000,-10){16}{\line(1,0){7}} 

\put(44,0){\circle*{3}}
\put(40,0){\line(1,0){8}}
\put(40,6){\framebox(8,8){X}}
\put(44,0){\line(0,1){6}}
\put(40,20){\line(1,0){8}}
\put(40,30){\line(1,0){8}}
\put(40,40){\line(1,0){8}}
\put(44,50){\circle*{3}}
\put(40,50){\line(1,0){8}}
\put(40,56){\framebox(8,8){X}}
\put(44,50){\line(0,1){6}}
\put(40,70){\line(1,0){8}}
\put(40,80){\line(1,0){8}}
\put(40,90){\line(1,0){8}}
\put(44,100){\circle*{3}}
\put(40,100){\line(1,0){8}}
\put(40,106){\framebox(8,8){X}}
\put(44,100){\line(0,1){6}}
\put(40,120){\line(1,0){8}}
\put(40,130){\line(1,0){8}}
\put(40,140){\line(1,0){8}}
\put(40,150){\line(1,0){8}}

\put(38,-6){\dashbox(12,122){}}

\multiput(48,150)(000,-10){16}{\line(1,0){7}} 

\put(55,-4){\framebox(8,8){Z}}
\put(55,6){\framebox(8,8){Y}}
\put(55,20){\line(1,0){8}}
\put(55,30){\line(1,0){8}}
\put(59,40){\circle*{3}}
\put(55,40){\line(1,0){8}}
\put(55,46){\framebox(8,8){Z}}
\put(59,46){\line(0,-1){32}}
\put(59,4){\line(0,1){2}}

\multiput(55,90)(0,-10){4}{\line(1,0){8}}

\put(55,96){\framebox(8,8){Z}}
\put(55,106){\framebox(8,8){Y}}
\put(55,120){\line(1,0){8}}
\put(55,130){\line(1,0){8}}
\put(59,140){\circle*{3}}
\put(55,140){\line(1,0){8}}
\put(55,150){\line(1,0){8}}
\put(59,140){\line(0,-1){26}}
\put(59,104){\line(0,1){2}}

\multiput(63,150)(000,-10){16}{\line(1,0){7}} 

\put(70,46){\framebox(8,8){Z}}
\put(70,56){\framebox(8,8){Y}}
\put(70,70){\line(1,0){8}}
\put(70,80){\line(1,0){8}}
\put(74,90){\circle*{3}}
\put(70,90){\line(1,0){8}}
\put(70,96){\framebox(8,8){Z}}
\multiput(70,40)(0,-10){5}{\line(1,0){8}}
\multiput(70,150)(0,-10){5}{\line(1,0){8}}
\put(74,96){\line(0,-1){32}}
\put(70,54){\line(0,1){2}}

\put(53,-6){\dashbox(27,152){}}

\multiput(78,150)(000,-10){16}{\line(1,0){7}} 

\put(85,-4){\framebox(8,8){Z}}
\put(85,6){\framebox(8,8){Y}}
\put(85,20){\line(1,0){8}}
\put(89,30){\circle*{3}}
\put(85,30){\line(1,0){8}}
\put(85,36){\framebox(8,8){Z}}
\put(89,36){\line(0,-1){22}}
\put(89,4){\line(0,1){2}}

\put(85,46){\framebox(8,8){Z}}
\put(85,56){\framebox(8,8){Y}}
\put(85,70){\line(1,0){8}}
\put(89,80){\circle*{3}}
\put(85,80){\line(1,0){8}}
\put(85,86){\framebox(8,8){Z}}
\put(89,86){\line(0,-1){22}}
\put(89,54){\line(0,1){2}}

\put(85,96){\framebox(8,8){Z}}
\put(85,106){\framebox(8,8){Y}}
\put(85,120){\line(1,0){8}}
\put(89,130){\circle*{3}}
\put(85,130){\line(1,0){8}}
\put(85,136){\framebox(8,8){Z}}
\put(85,150){\line(1,0){8}}
\put(89,136){\line(0,-1){22}}
\put(89,104){\line(0,1){2}}

\put(83,-6){\dashbox(12,152){}}

\multiput(93,150)(000,-10){16}{\line(1,0){7}} 

\put(100,-4){\framebox(8,8){Z}}
\put(100,6){\framebox(8,8){X}}
\put(104,20){\circle*{3}}
\put(100,20){\line(1,0){8}}
\put(100,26){\framebox(8,8){Z}}
\put(100,40){\line(1,0){8}}
\put(104,26){\line(0,-1){12}}
\put(104,4){\line(0,1){2}}

\put(100,46){\framebox(8,8){Z}}
\put(100,56){\framebox(8,8){X}}
\put(104,70){\circle*{3}}
\put(100,70){\line(1,0){8}}
\put(100,76){\framebox(8,8){Z}}
\put(100,90){\line(1,0){8}}
\put(104,76){\line(0,-1){12}}
\put(104,54){\line(0,1){2}}

\put(100,96){\framebox(8,8){Z}}
\put(100,106){\framebox(8,8){X}}
\put(104,120){\circle*{3}}
\put(100,120){\line(1,0){8}}
\put(100,126){\framebox(8,8){Z}}
\put(100,140){\line(1,0){8}}
\put(100,150){\line(1,0){8}}
\put(104,126){\line(0,-1){12}}
\put(104,104){\line(0,1){2}}

\put(98,-6){\dashbox(12,142){}}

\multiput(108,150)(000,-10){16}{\line(1,0){17}} 

\end{picture}
\caption[Encoding circuit for the 5-qubit convolutional code with the
pearl-necklace structure.]{Encoding circuit for the 5-qubit
convolutional code with the pearl-necklace structure. Each dashed box
represents a different layer in which the unitaries commute. Note that
the first three Hadamard gates cannot be put into a layer, but rather
form the unitary $I_{\mathrm{nit}}$.}\label{fig:pearl_necklace_qcc5}
\end{center}
\end{figure}
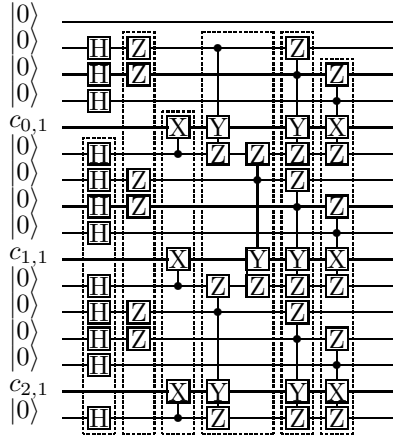

The pearl-necklace structure of the encoding circuit for the 5-qubit
convolutional code is presented in Fig.~\ref{fig:pearl_necklace_qcc5}.

\subsection{Catastrophicity condition for standard encoders}

\begin{proposition}
Encoders derived from the standard polynomial form are
non-catastrophic if and only if $\Lambda(D)$ is a monomial.
\end{proposition}

\begin{proof}\rm
Simple commutations rules between controlled gates can be used to show
that when $\Lambda(D)$ is a monomial, the quantum circuit can be put
in the form of Eq.~(\ref{eq:non_catastrophic}). To prove the
necessity, suppose $\Lambda(D)$ is not a monomial and consider the
decoding circuit for this code. More precisely, focus on the qubits
that control the application of $\overline X_{0,1}, \ldots \overline
X_{q-1,1}$. If the decoding circuit is restricted to those qubits
only, the only two-qubit gates that are used are
controlled-NOT's. Thus, this part of the quantum circuit in fact
implements a rate 1 classical convolutional encoder with
feedback. This encoder links its output stream $y(D)$ with its input
$x(D)$ through (see~\cite{Lee97a} for a rapid introduction to
classical convolutional codes and their polynomial formalism),
\begin{equation}
y(D) = x(D) + (\Lambda(1/D) - 1) y(D).
\end{equation}
Thus, an error affecting the input stream---corresponding to a bit
flip in the quantum case---propagates to an infinite number of
output bits when $\Lambda(D)$ is not a monomial: 
\begin{equation}
y(D) = \frac{x(D)}{\Lambda(1/D)}.
\end{equation}
Similarly, in the quantum case, a single bit flip could propagate to
an infinite number of qubits. Thus non-catastrophic standard encoders
have a monomial $\Lambda(D)$.\eend
\end{proof}

\begin{remark}\rm
Note also that the condition ``$\Lambda(D)$ is a monomial'' is
equivalent to having the $\overline Z$ operators efficiently described
with the polynomial formalism. These two questions are in fact
intimately related. The application of a $\overline Z$ can be done
before encoding by applying the corresponding $Z$ to the physical
unprotected qubit. It is well known that phase flips propagate through
controlled-NOT gates from the target to the control. Here, this phase
flip propagates in the same way the bit flip of the proof propagates
in the decoding circuit. The number of qubits affected by this $Z$
operation after running the encoding increases linearly with $q$, the
number of $k$-qubit blocks to be protected. More generally, the
non-catastrophicity condition shows that contrarily to classical
convolutional codes, an operation with finite support acting before
encoding cannot propagate to an infinite number of qubits after
encoding.\eend
\end{remark}

\section{Error estimation algorithm}\label{sec:correction}
The last subject that must be addressed to arrive at a theory of
quantum convolutional codes is the error estimation algorithm. A naive
attempt at finding the most likely error could be to search among all
the possible errors. In turn, this usually implies an exponential
complexity in the number of encoded qubits, thus making this scheme
impractical for large amounts of to-be-protected information. In this
section, a maximum likelihood estimation algorithm with a linear
complexity is provided. This algorithm is similar to its classical
analog, known as the Viterbi algorithm~\cite{Vit67a, Lee97a, JZ99a}.

\subsection{Notation}
To simplify the description of the algorithm, some additional notation
will be useful. Recall Eq.~(\ref{eq:general_structure}) which defines
the generators of the stabilizer group $M_{j,i}$. The expression
``block $j$'' will refer to the qubits involved in $M_{j,i}$ for $i =
1, \ldots, n-k$. The qubits are numbered in increasing number from
left to right, so that the first $m$ qubits and the last $n$ qubits of
the second block are those separated on
Eq.~(\ref{eq:general_structure}) by a dashed line. Note also that due
to the convolutional nature of the code and because of the definition
of $m$, the last $m$ qubits of block $j$ are the same as the first $m$
qubits of block $j+1$. The syndrome $s_{j,i}$ for a received stream of
information is the result of the projective measurement associated to
the $M_{j,i}$. It is equal to $+1$ (resp. $-1$) if the measured state
belongs to the $+1$ (resp. $-1$) eigenspace of $M_{j,i}$. An element
of the Pauli group of the transmitted qubits is said to be compatible
with the syndrome $s_{j,i}$ if it commutes (resp. anti-commutes) with
$M_{j,i}$ when $s_{j,i} = 1$ (resp. $-1$). An error candidate up to
block $j$ is an operator of the Pauli group defined on all the qubits
up to block $j$ and which satisfies all the syndromes up to block
$j$. The likelihood of an error candidate is the logarithm of the
probability of getting this particular error pattern according to the
channel model. Since we consider memoryless channels, the likelihood
is the sum of the logarithms of single-qubit-error probabilities.

\subsection{Quantum Viterbi algorithm}
The algorithm examines the syndromes block by block and updates a list
of error candidates among which one of them coincides with the most
likely error. All this algorithm is classical except the syndrome
extraction procedure. 

The value of the syndrome is obtained by the usual phase estimation
circuit: an ancillary qubit is prepared in the $\ket 0$ state;
undergoes a Hadamard gate; conditionally applies $M_{j,i}$; is once
again modified by a Hadamard gate; and is measured according to the
$Z$ observable. The result of this measure is the value of the
syndrome $s_{j,i}$.

\begin{algorithm}[Quantum Viterbi algorithm] 
~

{\tt Inputs}: (\,{\em i}) The list of syndromes $\{s_{j+1,i}\}_i$ for
$i=1,\ldots , n-k$; (\,{\em ii}) a list $\{E_j^{(e)}\}_{e}$ with $e
\in \{I,X,Y,Z\}^\otimes m$ of error candidates up to block $j$ such
that the element $E_j^{(e)}$ corresponding to the index $e$ has a
tensor product decomposition ending by $e$ for its last $m$ qubits,
and such that it maximizes the likelihood given the previous
constraint. The list $\{E_j^{(e)}\}_e$ is constructed recursively.

{\tt Step $j+1$}: For a given value of $e' \in \in
\{I,X,Y,Z\}^{\otimes m}$, consider all the possible $n$-qubit
extensions of the elements of $E_j^{(e)}$ such that:
\begin{itemize} 
\item they satisfy the syndromes $s_{j+1,i}$ for $i=1, \ldots,
n-k$;
\item they have the prescribed tensor product decomposition $e'$ on
their last $m$ positions.
\end{itemize}
By construction, these extensions are error candidates up to block
$j+1$. For each $e' \in \{I,X,Y,Z\}^{\otimes m}$ select one such
extension with maximum likelihood---take one at random among them in
case of tie. This constitutes the new list of error candidates
$\{E_{j+1}^{(e')}\}_{e'}$.

When all the syndromes have been taken care of in this way, select the
most likely error candidate of the list. This error candidate is one
of the most likely errors compatible with all the syndromes.
\end{algorithm}

\begin{proof}\rm
Consider a most likely error $E_p$ for the whole $p$ blocks of
syndromes. The truncation of this error to the first $p-1$ blocks,
$E_{p-1}$, is by construction an error candidate up to block
$p-1$. This error candidate has maximum likelihood given its
decomposition on the last $m$ qubits. If it was not the case, another
error candidate, $\tilde E_{p-1}$, with the same decomposition on the
last $m$ qubits could be extended up to block $p$ by concatenation
with the last $n$ Pauli operators of $E$. It would therefore have a
strictly greater likelihood than $E$. Recursively, this property holds
for $E_j$: it has maximum likelihood given its tensor product
decomposition on the last $m$ positions. Thus, at each step $j$ of the
algorithm, one element of the list coincides with the most likely
error up to block $j$.\eend
\end{proof}

\begin{remark}\rm
Note that in the encoding of quantum convolutional codes, we chose to
set to $\ket{0}$ some logical qubits that were not described by the
polynomial formalism. This was done formally by adding their
$\overline Z$ operators to the stabilizer group of the code. Hence
either the first and last steps of the algorithm should be modified to
take into account these extra syndromes.\eend
\end{remark}

\begin{remark}\rm
It is also important to understand that in the error estimation
algorithm presented above, the most likely error is known only at the
end of the algorithm. However, in practice the error candidates
considered at step $j$ all coincide except on the last few
blocks. Hence, the most likely error is known except on the last few
blocks. Some simulations for a depolarizing channel with error
probability less than 0.05 showed that keeping two blocks in the
5-qubit convolutional code was enough to estimate the most likely
error with high probability.\eend
\end{remark}

\section{Conclusion}
This article showed the basis of quantum convolutional coding. An
appropriate polynomial formalism has been introduced to handle the
codes efficiently and to make calculations consistently with their
specific structure. A procedure for deriving an encoding circuit with
linear gate complexity has been given together with a condition which
warrants the good behavior of this circuit with respect to error
propagation effects. Finally, the quantum Viterbi algorithm has been
given explicitly. This algorithm finds the most likely error with a
complexity growing linearly with the number of encoded qubits.

More importantly, as the reader familiar with classical convolutional
codes can notice, other error estimation algorithms, such as
Bahl's~\cite{BCJR74a} algorithm---a stepping stone toward
turbo-decoding---, can readily be employed with the codes described
here. Hence, quantum convolutional codes open a new range of efficient
error correction strategies.

This work was partially supported by ACI S\'ecurit\'e Informatique,
projet R\'eseaux Quantiques.


\end{document}